%% file: bipartivity.tex
\documentclass{article}

\usepackage[utf8]{inputenc}
\usepackage{amsmath}
\usepackage{amssymb}
\usepackage[pdftex]{hyperref}
\usepackage{graphicx}
\usepackage{grffile}
\usepackage{color}
\usepackage{booktabs}
\usepackage{subfigure}
\usepackage{multicol}

\definecolor{LinkColor}{rgb}{0, 0, 0}
\definecolor{UrlColor}{rgb}{0, 0, 0}
\definecolor{CiteColor}{rgb}{0, 0, 0}

\hypersetup{ 
    colorlinks=true, 
    pdfauthor={Jérôme Kunegis}, 
    citecolor=CiteColor, urlcolor=UrlColor, linkcolor=LinkColor
}

\newcommand{\wOnePointFive}{0.70\textwidth}
\newcommand{\wTwo}{0.47\textwidth}

\begin{document}

\title{
  Exploiting the Structure of Bipartite Graphs for Algebraic and Spectral
  Graph Theory Applications
}

\author{Jérôme Kunegis}

\maketitle

\begin{abstract}
  In this article, we extend several algebraic graph analysis methods to
  bipartite networks.  In various areas of science, engineering and
  commerce, many types of information can be represented as networks,
  and thus the discipline of network analysis plays an important role in
  these domains.  A powerful and widespread class of network analysis
  methods is based on algebraic graph theory, i.e., representing graphs
  as square adjacency matrices.  However, many networks are of a very
  specific form that clashes with that representation: They are
  bipartite.  That is, they consist of two node types, with each edge
  connecting a node of one type with a node of the other type. Examples
  of bipartite networks (also called \emph{two-mode networks}) are
  persons and the social groups they belong to, musical artists and the
  musical genres they play, and text documents and the words they
  contain. In fact, any type of feature that can be represented by a
  categorical variable can be interpreted as a bipartite network.
  Although bipartite networks are widespread, most literature in the
  area of network analysis focuses on unipartite networks, i.e., those
  networks with only a single type of node.  The purpose of this article
  is to extend a selection of important algebraic network analysis
  methods to bipartite networks, showing that many methods from
  algebraic graph theory can be applied to bipartite networks with only
  minor modifications.  We show methods for clustering, visualization
  and link prediction.  Additionally, we introduce new algebraic methods
  for measuring the bipartivity in near-bipartite graphs.
\end{abstract}

\section*{Introduction}
The term \emph{network analysis} refers to an area of research covering
the social sciences, computer science, economy, and others.  The
analysis of networks is central to sociology, in which the relations
between people are modeled as \emph{networks}, as well as to recent
studies of the World Wide Web, Web Science and others.  The emerging
field of \emph{Network Science} is solely dedicated to the study of
networks~-- owing its existence to the fact that a large number of
diverse data can be modeled as networks: Not only ties between people,
as in social networks, but also communication between people can be
modeled as a network in which each act of communication is a link
between two persons, transportation networks such as road, railroad and
airport networks in which nodes are places and links represent
transportation corridors, but also reference networks between
publications or pages on the Web.  Other examples are found in biology,
where interactions between metabolites form a metabolic network, and in
linguistics, where semantic relationships between words form a semantic
network.  Although they are taken from a highly diverse set of
application areas, these examples all have in common the underlying
model of a \emph{network}: a set of nodes connected by links.

The advantages of network analysis become clear once we consider the
breadth of mathematical tools available for analysing networks: From
simple numerical characterizations of networks such as the
\emph{clustering coefficient} and the \emph{diameter} to the analysis of
distributions associated with a network such as the \emph{degree
  distribution}, a single network analysis method can be applied to all
types of networks, giving insight to any type of network, even those not
envisioned by the developers of the original method.  Another important
class of network analysis methods is given by the specialized subfield
of \emph{algebraic graph theory}, in which graphs are analysed using
algebraic methods. Its main tool consists in representing a graph by a
matrix, and using matrix decompositions and other methods to derive
properties of the network.

\begin{figure}
  \centering
  \includegraphics[width=\wOnePointFive]{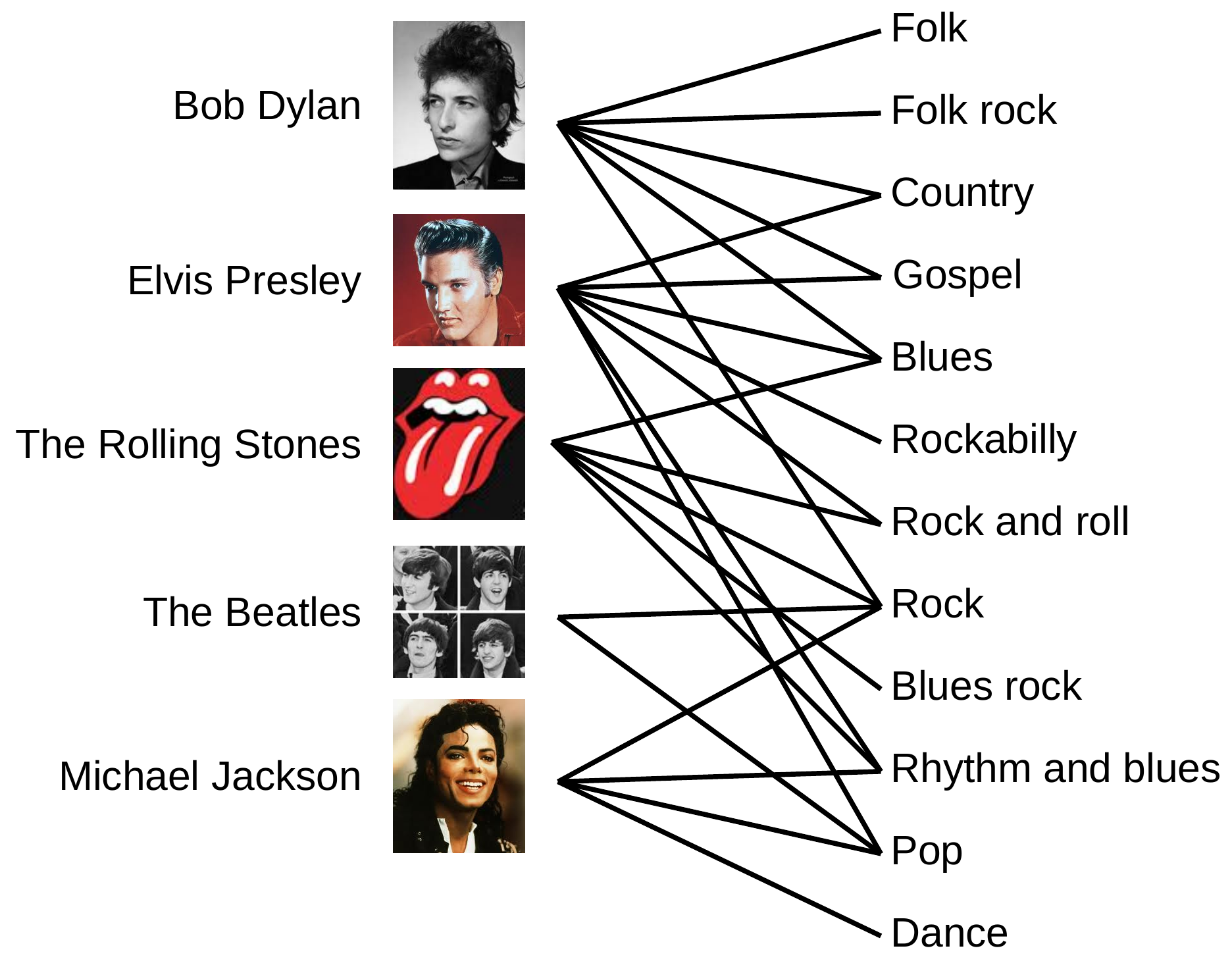}
  \caption{
    An example of a bipartite graph:  Musical artists and the genres
    they play. The network contains the two types of nodes \emph{artist}
    and \emph{genre}, and each edge connects a node of one type with a node
    of the other type.  
  }
  \label{fig:genres}
\end{figure}

Although networks are ubiquitous in many areas, networks are not all
similar: Many special types of networks exist, such as directed
networks, signed networks, weighted networks, and so on.  Although these
networks have differing definitions, network analysis methods for simple
networks can mostly be applied to them. For instance, instead of
defining the adjacency matrix (which we will define in detail later) as
a 0/1 matrix, it can be defined as an arbitrary matrix of reals, to
which the same methods can be applied as in the simple case.  One type
of network however is more complex in its structure: bipartite networks.
A bipartite network (also called \emph{two-mode network}) is a network
whose nodes can be partitioned into two sets, such that all links
connect two nodes of different types.  Ordinary networks such as social
networks for instance are not bipartite, since they contain triangles.
However, many other networks \emph{are} bipartite.  For instance, the
well-know \emph{Southern Women} dataset, consisting of information about
women's participation in social events in the Southern United States in
the 1930s, is bipartite~\cite{konect:southernwomen}.  For another
example, the set of connections between people and the things they like
forms a bipartite network, consisting of person-nodes and thing-nodes,
with each \emph{like} connecting a person with a thing.  Another example
are the countries of origin of persons or things, the musical genres
played by artists (as shown in Figure~\ref{fig:genres}), the teams in
which athletes have played, inclusion of people in social groups, tags
assigned to documents (or any other kind of resource) and ratings given
to movies (or other items).  Other, less obvious examples are the words
contained in text documents, in which the two node types are documents
and words, and each link denotes the \emph{contains} relationships. In
fact, bipartite networks are a very important class of networks:
Although the number cannot be taken as significant due to a clear
sampling bias, the Koblenz Network Collection~\cite{konect} contains, at
the time of writing, 42\% of bipartite networks (79 out of
189)\footnote{\href{http://konect.uni-koblenz.de/}{konect.uni-koblenz.de}}.
A sample of bipartite network types is given in
Table~\ref{tab:examples}.  A longer list is given in
Table~\ref{tab:networks} in Appendix~\ref{sec:networks}.

The artist--genre network shown in Figure~\ref{fig:genres} is a subset
of the actual Wikipedia genre information extracted by the DBpedia
project \cite{b642}, and will be used as a running example in the rest
of this article.  As a comparison, we will use a unipartite network
containing tie of members of a karate club, taken from a well-known
study by Wayne Zachary~\cite{konect:ucidata-zachary}. This small
unipartite network contains 34 nodes and 78 edges.

A large part of the network analysis literature and methods only apply
to unipartite networks, i.e., those networks having a single node type.
Therefore, many studies project bipartite networks to unipartite
networks, losing information in the process.  To avoid this, unipartite
algebraic graph analysis tools must be adapted, extended or completely
redefined in the bipartite case. This has been done partially for some
network analysis methods~\cite{b796}, although these methods do not
exploit algebraic graph theory.  This lack is the motivation of this
article: to present a selection of the most important algebraic network
analysis methods for bipartite networks.

\begin{table}
  \caption{
    \label{tab:examples}
    Examples of bipartite networks encountered in various areas.
  }
  \centering
  \begin{tabular}{lll}
    \toprule
    \textbf{Example} & \textbf{Node Types} & \textbf{References} \\
    \midrule
    Music genres & Artists + Genres & \cite{b642} \\
    Starring & Actors + Movies & \cite{b726} \\
    Sports & Players + Teams & \cite{b724} \\
    Authorship & Authors + Works & \cite{b222} \\
    Metabolism & Substances + Reactions & \cite{b720} \\
    Ratings & Users + Items & \cite{b520,b25,b343} \\
    Listening & Persons + Songs & \cite{lastfm} \\
    Affiliation & Persons + Groups & \cite{b560} \\
    Web analytics & Users + Web pages & \cite{b612} \\
    Search engines & Queries + Clicked URLs & \cite{b373} \\
    Economy & Producers + Consumers & \cite{b731} \\
    Tag assignment & Items + Tags & \cite{konect:bibsonomy} \\
    Bag of words & Documents + Words & \cite{b716} \\
    Taxonomy & Documents + Categories & \cite{b717} \\
    \bottomrule
  \end{tabular}
\end{table}

The outline of the article is as follows:
\begin{itemize}
\item \textbf{Section \ref{sec:definitions}, Bipartite graphs}:  We give
  the definition of a bipartite network and review alternative
  representation of bipartite datasets other than through networks. 
\item \textbf{Section \ref{sec:algebraic}, Algebraic graph theory}:  We
  review the usage of matrices  
  to represent networks, requiring the replacement of the adjacency
  matrix with the biadjacency matrix.  We show that paths between nodes
  in a bipartite network can be counted by computing alternating powers
  of the biadjacency matrix. 
\item \textbf{Section \ref{sec:measures}, Measures of bipartivity}:
  In the case where a network is not explicitly bipartite, it may be
  almost-bipartite. This section reviews algebraic measures of
  (non-)bipartivity, introducing several new ones.  
\item \textbf{Section \ref{sec:clustering}, Clustering analysis}: This
  section describes spectral methods for 
  detecting and measuring clustering in a network, i.e., the tendency of
  edges to form tight groups, as well as methods for finding such
  clusters. 
\item \textbf{Section \ref{sec:visualization}, Visualization}:  We
  describe methods for visualizing a 
  bipartite network, showing that the bipartite structure is an
  additional type of information that can or cannot
  be visualized, reviewing advantages of both variants. 
\item \textbf{Section \ref{sec:link-prediction}, Link prediction}:  We
  treat the problem of link prediction, i.e., the prediction of links
  in evolving networks and show how common link prediction graph can be
  generalized to the bipartite case. 
\item \textbf{Section \ref{sec:other}, Other graph types}:  We review
  other graph types such as weighted and signed graphs, and their
  application to the bipartite case. 
\end{itemize}

Throughout the paper, examples of bipartite networks will be taken from
the Koblenz Network Collection~\cite{konect}, a collection of
small and large networks of different types and from various application
areas created by the authors. 
Section~\ref{sec:link-prediction} (Link Prediction) is partly based on
a previous paper of the author~\cite{kunegis:hyperbolic-sine}. 

\section{Bipartite Graphs}
\label{sec:definitions}
A network is an abstract structure consisting of nodes connected by
links. Networks are modeled mathematically as graphs, which consist of a
set of vertices (representing the nodes) connected by a set of edges
(representing the links).  
A graph is usually denoted as
\begin{align*}
  G = (V, E),
\end{align*}
in which $G$ is the graph, $V$ is the set of vertices, and $E$
is the set of edges.  An edge connecting the vertices $u, v \in
V$ is denoted $\{u,v\} \in E$. 
The fact that two vertices $u$ and $v$ are connected will also be
denoted $u \sim v$. The number of neighbors of a node $u$ is called the
degree of $u$ and is denoted $d(u)=\{v \in V \mid u \sim v\}$. 

Bipartite graphs are graphs in which the set of nodes
can be partitioned into two disjoint sets such that each edge connects
a vertex of one partition with a vertex of the other partition.  A
bipartite graph can be denoted as
\begin{align*}
  G &= (V_1, V_2, E),
\end{align*}
in which $V_1$ and $V_2$ are the two disjoint vertex sets and $E$ is the set
of vertices. 

An equivalent characterization of bipartite graphs are the graphs that
do not contain odd cycles, i.e., cycles consisting of an odd number of
edges. In particular, it follows that a bipartite network does not
contain triangles. 

The term \emph{bipartite} is sometimes employed in a slightly different
way in mathematics and network science. 
Mathematically, any graph $G=(V,E)$ is by definition
bipartite when a bipartition of $V$ exists, such that all edges in $E$
connect a vertex of one partition with a vertex of the other
partition. In our notation however, a bipartition of the vertices will
always be explicitly given by using the notation $G=(V_1, V_2, E)$.
Thus, the term \emph{unipartite} is not the exact opposite of
\emph{bipartite}:  We call a network unipartite when it has a single
node type, and links can connect any two nodes. This definition does not
preclude a unipartite network from being bipartite in the mathematical
sense. In the rest of the article, we will restrict usage of the term
\emph{bipartite} and \emph{unipartite} to networks with two and one
explicit node types\footnote{Another terminology, \emph{1-mode network}
and \emph{2-mode network}, is also common.}.  
The two node sets $V_1$ and $V_2$ will be called the left and right edge
sets\footnote{Other designations exist; for instance \emph{top} and
  \emph{bottom}~\cite{b725}.}.   

Although bipartivity gives a special structure to a network, which can
be exploited in various ways, we are not interested here in graphs that
cannot be described as \emph{complex networks}. For instance, a forest
(i.e., a cycle-free graph) is bipartite in mathematical sense, but its
structure is too trivial to be interesting in most cases. 

\subsection{Alternative Representations}
An alternative representation of bipartite networks is as
hypergraphs~\cite{b364}. 
Let $G=(V_1, V_2, E)$ be a bipartite graph. Then we can construct its
derived hypergraph $\mathrm{Hyp}(G)=(V_1, F)$. As a reminder, the notion
of hypergraph extends the notion of graph by allowing edges to
connect any number of vertices, and are usually called
\emph{hyperedges}.  By setting 
\begin{align*}
F= (\{v \in V_2 \mid \{u,v\}\in E\})_{u \in V_1},
\end{align*}
we arrive at a hypergraph $H$
in which the left vertices of $G$ are the vertices, and the neighbors of
each left node $u$ in $V_1$ make one hyperedge in $H$. In reverse, each
hypergraph can be reduced to a bipartite graph in this way~-- another
reason why bipartite graphs are important. As an example, a sets of
groups formed by persons on a social networking site can be modeled as a
hypergraph, in which the persons are the nodes, and each group gives one
hyperedge making up all persons contained in that group. The equivalent
bipartite graph has persons and groups as nodes, and an edge from a
person to a group when the person is member of that group. 
Due to the symmetry between $V_1$ and $V_2$, we can build an analogous
hypergraph representation $\mathrm{Hyp}(G)=(V_2, F)$ in which $F$
contains one hyperedge for each node in $V_1$. 

Another alternative way to model bipartite data is using the vector
space model. This approach is very
common in text mining, where documents containing words are
modeled as word-vectors. 
Let $G=(V_1, V_2, E)$ be a bipartite document--word network, where $V_1$
are the documents and $V_2$ are the words. The vector space model then
represents each document $u \in V_1$ as a vector $\mathbf x^u \in \mathbb
R^{|V_2|}$ defined as
\begin{align*}
  [\mathbf x^u]_v &= \left\{ \begin{array}{cc}
    1 & \text{ when } \{u,v\} \in E \\
    0 & \text{ when } \{u,v\} \notin E 
  \end{array} \right.
\end{align*}
In this representation, certain measures arise naturally, such as the
cosine similarity. However, more complex methods are harder to
recover. As an example, one may consider the cosine similarity, which
measures the cosine of the angle between two vectors, or equivalently
the dot product of two vectors, after a suitable normalization.  By
construction, the cosine similarity takes into account common words
contained in two documents, but not more complex semantic relationships,
such as synonyms. Instead, a graph-based similarity measure that
considers paths in the bipartite document--word network will be able to
find such relationships, as long as it is based on paths of length more
than two between two documents. 

Another common alternative representation of a bipartite network is
by its projection to a unipartite network~\cite{b732}. 
In the projection of a network onto the left nodes, only the left nodes
are kept, and two nodes are connected when they have a common neighbor
in the original bipartite graph.  The projection onto the right nodes is
defined analogously. 
Let $G=(V_1,V_2,E)$ be a bipartite graph.  Then its projections to the
left and right side can be defined as the graphs
\begin{align*}
  G_{\mathrm L} &= (V_1, \{ \{u,v\} \mid \exists w \in V_2: u \sim w, v \sim w \}) \\
  G_{\mathrm R} &= (V_1, \{ \{u,v\} \mid \exists w \in V_1: u \sim w, v \sim w \}) 
\end{align*}
The projections defined in this way are commonly used when unipartite
network analysis methods are to be applied to bipartite networks. 
Among many examples, this is the case for edge types
representing collaboration, 
whose name typically begin with \emph{co-}, for instance, the co-authorship network of
scientists or the co-starring network of actors. 
The projection networks however do not fully reflect the properties of
the original bipartite networks. For instance, the left and right degree
distributions in the original bipartite graph will by combined.
However, the left and right
degree distributions of bipartite networks are often very different, and
this is lost in the projection. 

\subsection{Size, Volume and Fill}
\begin{figure}
  \centering
  \includegraphics[width=0.6\textwidth]{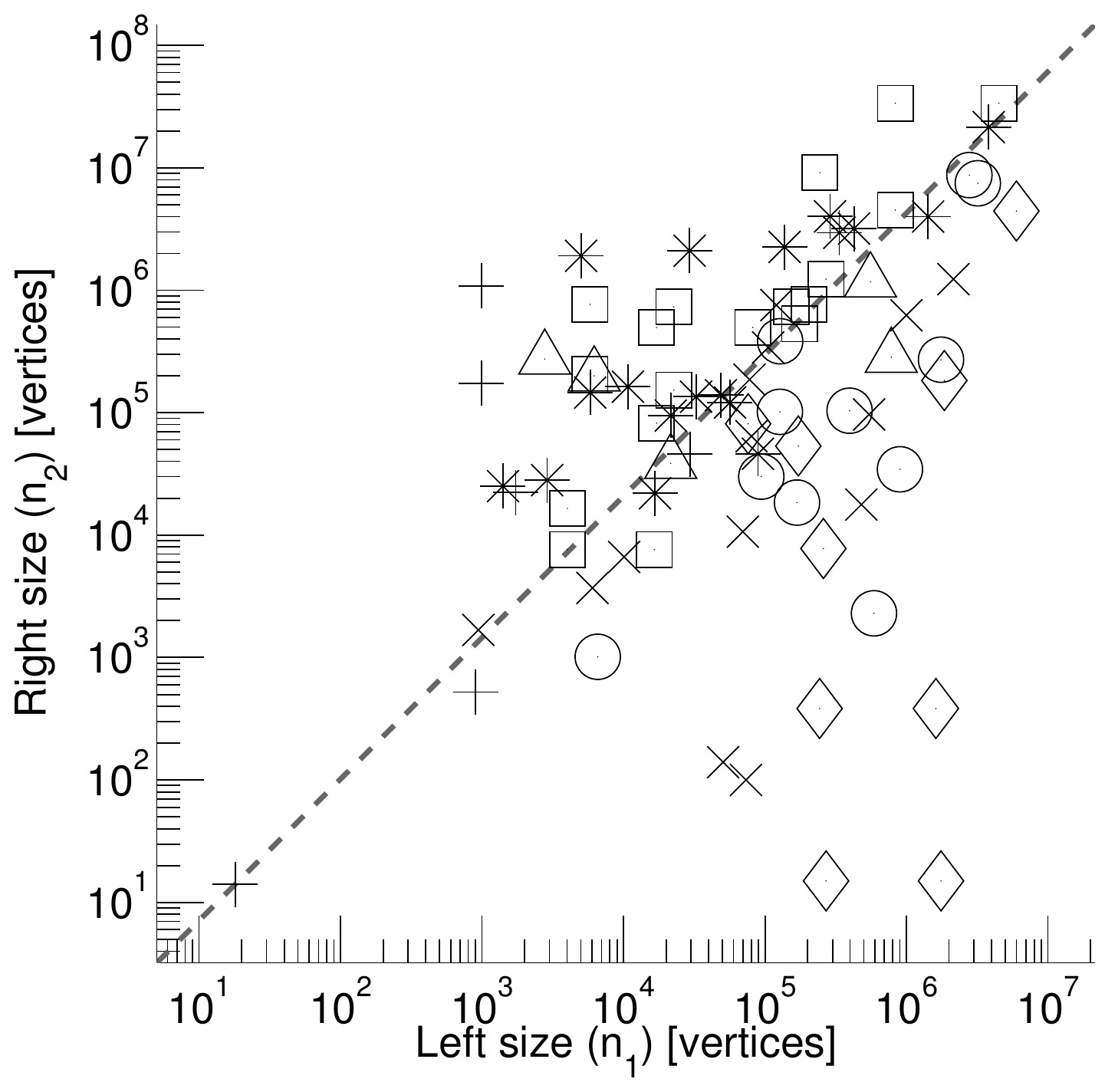}
  \raisebox{3cm}{
    \includegraphics[width=0.28\textwidth]{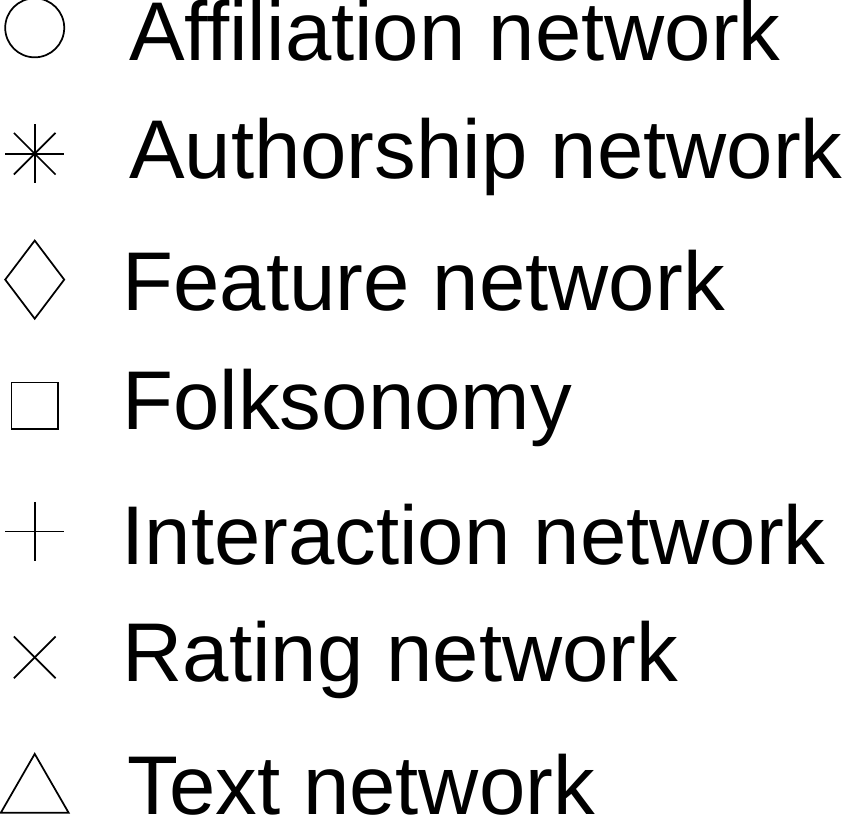}}
  \caption{
    Comparison of the left and right network sizes (number of nodes) in
    bipartite networks of the Koblenz Network Collection~\cite{konect}. 
    Each letter code stands for one bipartite network. 
  }
  \label{fig:scatter.size-2.size-3}
\end{figure}

The size, volume and fill are three basic network characteristics that
extend trivially to bipartite networks. 
The size denotes the number of nodes in a network.  The size of a graph
$G=(V,E)$ is $|V|$.  For a bipartite graph $G=(V_1, V_2, E)$, the size is
$|V_1|+|V_2|$, and we can distinguish between the left size $|V_1|$ and the
right size $|V_2|$. 
The sizes of the left and right node sets can vary
greatly. Figure~\ref{fig:scatter.size-2.size-3} shows a scatter plot of
the left and right node sets of all networks in the Koblenz Network
Collection~\cite{konect}. As 
an example 
for a very skewed bipartite network, the rating network of Netflix~\cite{b520}
contains 480,189 users but only 17,770 movies. 

The volume of a graph $G=(V,E)$ is its number of edges.  For a bipartite
graph $G=(V_1, V_2,E)$, no change in the definition is needed; the volume
is simply $|E|$.  

The fill of a network is the proportion of edges to the total amount of
possible edges. In a unipartite graph $G=(V,E)$, the fill is given by
$f = |E| / {|V| \choose 2}$. In a bipartite graph $G=(V_1, V_2 ,E)$, only edges between nodes
in $V_1$ and nodes in $V_2$ are allowed, and thus the fill is given by
$f=|E|/(|V_1||V_2|)$. 

\section{Algebraic Graph Theory}
\label{sec:algebraic}
Algebraic graph theory is the branch of graph theory that represents
graphs using algebraic structures in order to exploit the 
methods of algebra for graph theory. The main tool of algebraic graph
theory is the representation of graphs as matrices, in particular the
adjacency matrix and the Laplacian matrix. In the scope of this article,
we will look at the adjacency matrix, since it has a special structure
in the bipartite case, which can be exploited to give insights about
path counts in the network. 
In the following, all matrices are real. 

\subsection{Adjacency Matrix}
Given a unipartite graph $G=(V,E)$, its adjacency matrix $\mathbf A \in \mathbb
R^{|V|\times |V|}$ is defined by
\begin{align*}
  \mathbf A_{uv} &= \left\{ \begin{array}{lll} 1 & \text{when } \{u,v\}
    \in E \\ 0 & \text{when } \{u,v\} \notin E \end{array} \right. 
\end{align*}
The adjacency matrix is square and symmetric,
and has the useful property that it can be used to
count paths in the network. It is easy to verify that the entry of the
square of the adjacency matrix $[\mathbf A^2]_{uv}$ equals the number of common neighbors
between the vertices $u$ and $v$, or equivalently the number of paths of length
two between them. This result can be generalized to any power. For any
$k \geq 0$, the
number $[\mathbf A^k]_{uv}$ equals the number of paths of length $k$
between the nodes $u$ and $v$. 
Counting the number of paths between two nodes is a very useful tool in
network analysis.  For instance, it can be used to recommend new friends
in a social network (the \emph{friend of a friend} approach), or to
compute the distances from one node to all other nodes. 

Due to its structure, a bipartite graph $G = (V_1, V_2, E)$ has an
adjacency matrix of the form
\begin{align}
  \mathbf A &= \left[ \begin{array}{cc} \mathbf 0 & \mathbf B \\ \mathbf
      B^{\mathrm T} & \mathbf 0 \end{array} \right],
  \label{eq:matrix-structure}
\end{align}
in which $\mathbf B \in \mathbb R^{|V_1|\times |V_2|}$ is called the
biadjacency matrix of $G$, and $\mathbf 0$ are zero matrices of
appropriate size. 
Because the biadjacency matrix is rectangular, we cannot simply take its
powers to count paths in the network.  
However, we can derive a specific form of powers 
from its relation to the adjacency matrix, making a distinction between
even and odd powers: 
\begin{align*}
  \mathbf A^{2k} &= \left[ \begin{array}{cc} 
      (\mathbf B \mathbf B^{\mathrm T})^k & \mathbf 0 \\
      \mathbf 0 & (\mathbf B^{\mathrm T} \mathbf B)^k 
    \end{array} \right], \\
  \mathbf A^{2k+1} &= \left[ \begin{array}{cc} 
      \mathbf 0 & (\mathbf B \mathbf B^{\mathrm T})^k \mathbf B  \\
      (\mathbf B^{\mathrm T} \mathbf B)^k \mathbf B^{\mathrm T} & \mathbf 0 
    \end{array} \right]. \\
\end{align*}
From this we can derive expressions for path counts: 
\begin{itemize}
\item The number of paths of length $2k$ between two nodes $u,v \in V_1$
  equals $[(\mathbf B \mathbf B^{\mathrm T})^k]_{uv}$. 
\item The number of paths of length $2k$ between two nodes $u,v \in V_2$
  equals $[(\mathbf B^{\mathrm T} \mathbf B)^k]_{uv}$.
\item The number of paths of length $2k+1$ between a node $u\in V_1$ and
  a node $v\in V_2$ is $[(\mathbf B \mathbf B^{\mathrm T})^k \mathbf
  B]_{uv}$.
\end{itemize}
All other path counts are zero due to the bipartite structure of the
graph. 
Thus, all path counts in a bipartite graph can be computed using
alternating powers of the biadjacency matrix, following the pattern
$\mathbf B \mathbf B^{\mathrm T}  \dotsm  \mathbf B$. 

\subsection{Matrix Decompositions}
A major advantage of representing networks as matrices is the
possibility of exploiting matrix decompositions, which can be used to
compute powers of matrices efficiently. 
In particular, two decompositions are used:  the eigenvalue decomposition
and the singular value decomposition.  

Given a real symmetric matrix\footnote{The eigenvalue decomposition is
  also defined for more general matrices, but these cases are not needed
in this article.} $\mathbf X \in \mathbb R^{n \times n}$,
its eigenvalue decomposition is
\begin{align*}
  \mathbf X &= \mathbf U \mathbf \Lambda \mathbf U^{\mathrm T},
\end{align*}
in which $\mathbf U$ is an orthogonal matrix of size $n \times n$ (meaning that $\mathbf
U^{\mathrm T} \mathbf U = \mathbf I$) and $\mathbf \Lambda$ is a real diagonal matrix of
size $n \times n$. 
The diagonal elements of $\mathbf \Lambda$ are called the eigenvalues of
$\mathbf X$, and the columns of $\mathbf U$ are called its
eigenvectors. 
The set of eigenvalues of a matrix are also called its spectrum. 
The multiplicity of an eigenvalue is defined as the number of times it
occurs on the diagonal of $\mathbf \Lambda$. 

Given any $m \times n$ real matrix $\mathbf Y$ (square or non-square, symmetric or asymmetric),
its singular value decomposition is
\begin{align*}
  \mathbf Y &= \mathbf U \mathbf \Sigma \mathbf V^{\mathrm T},
\end{align*}
in which $\mathbf U$ and $\mathbf V$ are orthogonal matrices of sizes
$m\times m$ and $n \times n$, and $\mathbf \Sigma$ is a nonnegative diagonal $m \times n$
matrix. The diagonality of $\mathbf \Sigma$ is to be understood as
\emph{all entries $\mathbf \Sigma_{ij}$ with $i \neq j$ are zero}. 
The diagonal elements of $\mathbf \Sigma$ are called the singular values
of $\mathbf Y$; the columns of $\mathbf U$ and $\mathbf V$ are called
its left and right singular vectors. 

The eigenvalue decomposition of a bipartite network's adjacency matrix
$\mathbf A$ is 
equivalent to the 
singular value decomposition of its biadjacency matrix $\mathbf B$. Given the singular value
decomposition $\mathbf B=\mathbf U \mathbf \Sigma \mathbf
V^{\mathrm T}$, the eigenvalue decomposition of $\mathbf A$ is given by
\begin{align*}
  \mathbf A &=
  \left[ \begin{array}{cc} \mathbf {\bar U} & {\phantom -}\mathbf
      {\bar U} \\ \mathbf {\bar V} & -\mathbf {\bar V} \end{array} \right] 
  \left[ \begin{array}{cc} +\mathbf \Sigma &  \\ &
      -\mathbf \Sigma \end{array} \right] 
  \left[ \begin{array}{cc} \mathbf {\bar U} & {\phantom -}\mathbf
      {\bar U} \\ \mathbf {\bar V} & -\mathbf {\bar V} \end{array}
    \right]^{\mathrm T},
\end{align*}
with $\mathbf {\bar U} =\mathbf U/\sqrt 2$ and $\mathbf {\bar V} =
\mathbf V/\sqrt 2$.
In this decomposition, 
each singular value $\sigma$ corresponds to the eigenvalue pair $\{\pm
\sigma\}$.  

In order to compute powers of the adjacency matrix, a special property
of the eigenvalue decomposition can be exploited. Using the eigenvalue
decomposition
$\mathbf A = \mathbf U \mathbf \Lambda \mathbf U^{\mathrm T}$, any power
of the adjacency matrix can be computed as
\begin{align*}
  \mathbf A^k &= \mathbf U \mathbf \Lambda^k \mathbf U^{\mathrm T}.
\end{align*}
This can be shown by using the fact the $\mathbf U$ is orthogonal, for instance,
$\mathbf A^2 = \mathbf U \mathbf \Lambda \mathbf U^{\mathrm T} \mathbf U
\mathbf \Lambda \mathbf U^{\mathrm T} = \mathbf U \mathbf \Lambda
\mathbf \Lambda \mathbf U^{\mathrm T}$. This form is convenient because
powers of the diagonal matrix $\mathbf \Lambda$ can simply be computed by
taking powers of its diagonal entries. 

To compute alternating powers of the biadjacency matrix $\mathbf B$, its
singular value decomposition can then be used in an analogous way.  For instance,
\begin{align*}
  \mathbf B \mathbf B^{\mathrm T} \mathbf B &= \mathbf U \mathbf \Sigma
  \mathbf V^{\mathrm T} \mathbf V \mathbf \Sigma \mathbf U^{\mathrm T}
  \mathbf U \mathbf \Sigma \mathbf V^{\mathrm T} \\
  &= \mathbf U \mathbf \Sigma^3 \mathbf V^{\mathrm T}. 
\end{align*}
This method can be used to compute any odd alternating power of $\mathbf
B$.  

\subsection{Graph Laplacian}
Beyond the adjacency matrix $\mathbf A$, another matrix is
commonly used to analyse graphs algebraically:  
the Laplacian matrix $\mathbf L$. 
Let $G=(V,E)$ be a graph with adjacency matrix $\mathbf A$.  Then, 
its degree matrix $\mathbf D$ is defined as the diagonal matrix
containing the node degrees on the diagonal, i.e.
\begin{align*}
  \mathbf D_{uu} &= d(u).
\end{align*}
The Laplacian matrix $\mathbf L$ is then the symmetric $|V|\times|V|$ matrix
defined as $\mathbf L = \mathbf D - \mathbf A$.  The Laplacian matrix
has only nonnegative eigenvalues (it is positive-semidefinite), and its
eigenvectors can be used for grouping the nodes of the graph into
clusters, and also for graph drawing, as we will see in sections
\ref{sec:clustering} and \ref{sec:visualization}. 

Unlike the adjacency matrix, the Laplacian matrix of a bipartite graph
does not have a block structure akin to that given in
Equation~\ref{eq:matrix-structure}.  Thus, we cannot define any
corresponding rectangular \emph{bi-Laplacian matrix} of which the
singular value decomposition would give us information equivalent to the
eigenvalue decomposition of the Laplacian $\mathbf L$~\cite{b377}. 

\section{Measures of Bipartivity}
\label{sec:measures}
Most bipartite networks are explicitly stored as such in information
systems. For instance, any recommender system will have a clear
distinction between \emph{users} and \emph{items}.  In these cases, it
is clear how to split the vertex set $V$ into $V_1$ and $V_2$. In other
cases, the bipartition is not explicit, but implicit in the data. As an
example, a dating website\footnote{For the sake of this example, we
  assume a heterosexual majority of users.} in which users can rate the
profile of other users will have a majority of edges connecting men with
women, and thus not be bipartite, but nearly
so~\cite{kunegis:split-complex-dating}. In this case, a bipartition
would have to be inferred from the data.  If the network is strictly
bipartite, a bipartition can then be derived easily. If however, as is
likely in the example of the dating website, the network is not quite
bipartite, then that is not possible.

Almost-bipartite networks include networks of sexual
contacts~\cite{b719} and ratings on online dating
sites~\cite{b311,kunegis:split-complex-dating}.  Other, more subtle
cases, involve online social networks.  For instance, the follower graph
of the microblogging service Twitter is by construction unipartite, but
has been observed to reflect, to a large extent, the usage of Twitter as
a news service~\cite{b545}, as two types of users can be identified:
those who primarily get followed (news sources) and those who primarily
follow (readers).  Thus, the Twitter follower graph is almost bipartite.
Other social networks do not necessarily have a near-bipartite
structure, but the question might be interesting to ask to what extent a
network is bipartite.  To answer this question, we need to define
measures of bipartivity.

Instead of defining measures of bipartivity, we will instead consider
measures of non-bipartivity, as these can be defined in a way that they
equal zero when the graph is bipartite.  Given an (a priori) unipartite
graph $G=(V,E)$, a measure of non-bipartivity $b(G)$ characterizes the
extent to which $G$ fails to be bipartite.  In the following, we review
four spectral measures of non-bipartivity, and compare them numerically
on a large collection of example datasets.  All four measures we present
are nonnegative, and equal zero if and only if the graph is
bipartite.\footnote{ We note that the first measure of non-bipartivity
  is presented in its reference as a measure of bipartivity to which we
  applied the transformation $x \rightarrow 1-x$.  } The four measures
are based on the three characteristic graph matrices $\mathbf A$,
$\mathbf N$ and $\mathbf K$. Two of the measures are novel, and one is a
novel algebraic relaxation of a known non-algebraic measure.  All four
measures are algebraic. Incidentally, we are not aware of a
non-algebraic measure of (non-)bipartivity that is not covered by our
four measures.

\subsection{Frustration -- Spectrum of $\mathbf K$}
A first test of bipartivity consists in counting the minimum number of
\emph{frustrated edges} in a network~\cite{b531,b797}. Given a
bipartition $V=V_1\cup V_2$, a frustrated edge is an edge connecting two
nodes in $V_1$, or two nodes in $V_2$.  Let $f$ be the minimal number of
frustrated edges in any bipartition of $V$. A measure of non-bipartivity
is then given by the ratio of frustrated edges to total edges
\begin{align*}
  b_{\mathrm f}(G) &= \frac f {|E|}.
\end{align*}
This measure is always in the range $[0, 1/2]$.  It attains the value
zero if and only if $G$ is bipartite.  The number of possible partitions
$V=V_1 \cup V_2$ in this minimization problem is exponential in the
number of nodes, and thus the problem cannot be solved easily. Instead,
it can be solved approximately using a Monte-Carlo algorithm.
Alternatively, the minimal number of frustrated edges $f$ can be
approximated by algebraic graph theory, in a way which we introduce
below.  First, we represent a bipartition $V = V_1 \cup V_2$ by its
characteristic node-vector $\mathbf x \in \mathbb{R}^{|V|}$ defined as
\begin{align*}
  \mathbf x_u &= \left\{ \begin{array}{ll}
    +1/2 & \text{when $u \in V_1$} \\
    -1/2 & \text{when $u \in V_2$}
    \end{array} \right. 
\end{align*}
Note that the frustration $f$, i.e., the number of edges within the sets
$V_1$ and $V_2$ is then given by
\begin{align*}
f &=
\frac 12 \sum_{(u,v) \in E} (\mathbf x_u + \mathbf x_v)^2
= \frac 12 \mathbf x^{\mathrm T} \mathbf K \mathbf x,
\end{align*}
where $\mathbf K = \mathbf D + \mathbf A$ is the
signless Laplacian matrix of the underlying unweighted graph. 
Thus, the minimal number of frustrated edges $f$ is given by
\begin{align*}
  f &=  \min_{\mathbf x \in \{\pm 1/2\}^{|V|}}
  \frac 12 \mathbf x^{\mathrm T} \mathbf K \mathbf x. 
\end{align*}
By relaxing the condition $\mathbf x \in \{\pm 1/2\}^{|V|}$, we can
express $f$ in function of $\mathbf K$'s minimal eigenvalue, using the
fact that the norm of all vectors $\mathbf x \in \{\pm 1/2\}^{|V|}$
equals $\sqrt{|V|/4}$, and the property that the minimal eigenvalue of a
matrix equals its minimal Rayleigh quotient.
\begin{align*}
  \frac {2f} {|V|/4}
  &\approx \min_{\mathbf x \neq \mathbf 0} \frac {\mathbf x^{\mathrm T} \mathbf K
    \mathbf x} {\left\| \mathbf x \right\|^2}  
  = \lambda_{\min}[\mathbf K]  
\end{align*}
We can thus approximate the measure of non-bipartivity $b_{\mathrm f}$
by
\begin{align*}
  b_{\mathrm f} &\approx b_{\mathrm K} = \frac {|V|} {8 |E|} 
  \lambda_{\min}[\mathbf K] 
\end{align*}
Finally, we note that the eigenvalue $\lambda_{\min}[\mathbf K]$ can also be
interpreted as the algebraic conflict in $G$ interpreted as a signed
graph in which all edges have negative weight~\cite{kunegis:negativity}. 

\subsection{Spectrum of $\mathbf A$}
The adjacency matrix $\mathbf A$ is symmetric and thus has real
eigenvalues.  In general, its eigenvalues are unbounded and can be
positive and negative.  When a network does not contain loops (edges
from a node to itself), the trace of $\mathbf A$ and thus the sum of its
eigenvalues is zero. However, the distribution of eigenvalues around
zero is not symmetric in the general case.  In fact, the eigenvalues are
distributed in a symmetric way around zero if and only if the graph is
bipartite.  Thus, the eigenvalues of $\mathbf A$ can serve as a test of
non-bipartivity in the following way: When a graph is bipartite, the
smallest eigenvalue of the adjacency matrix equals the negative of the
largest one, i.e., $\lambda_{\min}[\mathbf A] = - \lambda_{\max}[\mathbf
  A]$. If a graph is not bipartite, the smallest eigenvalue is nearer to
zero, i.e., $\lambda_{\min}[\mathbf A] > - \lambda_{\max}[\mathbf A]$.
Note that it is not possible that the smallest eigenvalue equals the
largest in absolute value in a non-bipartite graph.  Thus, the ratio of
the smallest eigenvalue to its minimal value as a measure of
non-bipartivity, which, when scaled accordingly, is zero if and only if
the graph is bipartite.
\begin{align*}
  b_{\mathrm A}(G) &= 1 - \left| \frac
      {\lambda_{\min}[\mathbf A]} 
      {\lambda_{\max}[\mathbf A]} 
      \right| 
\end{align*}
This measure of non-bipartivity is always in the range $[0,1]$.  

\subsection{Spectrum of $\mathbf N$}
Instead of measuring the non-bipartivity of a graph by the eigenvalues
of its adjacency matrix $\mathbf A$, we can measure it by the
eigenvalues of a normalized version of the adjacency matrix, the matrix
$\mathbf N$.  The normalized adjacency matrix $\mathbf N$ has the same
structure as the adjacency matrix, but its entries are normalized by the
node degrees.  It is given by
\begin{align*}
  \mathbf N &= \mathbf D^{-1/2} \mathbf A \mathbf D^{-1/2}
\end{align*}
or equivalently by
\begin{align*}
  \mathbf N_{uv} &= \left\{ \begin{array}{ll} 
    1/ \sqrt{d(u)d(v)} & \text{ when $u$ and $v$ are connected} \\
    0  & \text{ when $u$ and $v$ are not connected}
    \end{array} \right. 
\end{align*}
The normalized adjacency matrix $\mathbf N$ can replace the adjacency
matrix $\mathbf A$ when a normalization of edge weights by degree is
appropriate. 

Due to the normalization, the eigenvalues of $\mathbf N$ have special
properties.  All eigenvalues of $\mathbf N$ are contained in the
interval $[-1,+1]$, the maximum eigenvalue $\lambda_{\max}[\mathbf N]$
is always one, and the minimal eigenvalue $\lambda_{\min}[\mathbf N]$
equals negative one if and only if the network is bipartite. Thus, we
propose as a measure of non-bipartivity
\begin{align*}
  b_{\mathrm N}(G) &= \lambda_{\min}[\mathbf N] + 1.
\end{align*}
This measure is always in the range $[0,1]$.

\subsection{Odd Cycles}
Another test for bipartivity consists in counting the cycles in the
network. If a network is almost bipartite, there are much more cycles of
even length than cycles of odd length. Thus, the proportion of odd
cycles to the total number of cycles, appropriately weighted to avoid
infinite cycle counts, can be used as a measure of
non-bipartivity~\cite{b456}. 

In this derivation, we consider cycles that may contain repeated
nodes\footnote{When cycles are defined to not contain any node twice,
  cycles as defined here may be called tours.}.  A cycle of length $k$
is a sequence of $k$ nodes $(u_1, u_2, \ldots, u_k)$ such that $u_i \sim
u_{i+1}$ for $1 \leq i < k$ and $u_1 \sim u_k$.  We thus consider cycles
to have a distinct starting node and a direction.  Given a graph $G$,
let $P(k)$ be the number of cycles of length $k$ in $G$.  Using the
adjacency matrix $\mathbf A$ of $G$, we can express this as
\begin{align*}
  P(k) &= \mathrm{Tr}(\mathbf A^k)
\end{align*}
Then, a measure of non-bipartivity is given by the ratio of odd cycles to all
cycles
\begin{align*}
  b_{\mathrm c}(G) &= \frac 
  {\sum_{k=0}^\infty w(1+2k) P(1+2k)}
  {\sum_{k=0}^\infty w(k) P(k)} \\
  &= \frac
  {\sum_{k=0}^\infty w(1+2k) \mathrm{Tr}(\mathbf A^{1+2k})}
  {\sum_{k=0}^\infty w(k) \mathrm{Tr}(\mathbf A^{k})} \\
  &= \frac
  {\mathrm{Tr}(\sum_{k=0}^\infty w(1+2k) \mathbf A^{1+2k})}
  {\mathrm{Tr}(\sum_{k=0}^\infty w(k) \mathbf A^{k})} \\
\end{align*}
This measure is dependent on a length-dependent cycle weight $w(k)$,
without which both sums would diverge.  Several choices for $w(k)$ are
possible.  We will consider here the weights $w(k) = 1/k!$, resulting in
the matrix exponential and hyperbolic sine:
\begin{align*}
  b_{\mathrm c} (G)
  &= \frac
  {\mathrm{Tr}(\sinh(\mathbf A))}
  {\mathrm{Tr}(\exp (\mathbf A))}
\end{align*}
This can be expressed as a function of $\mathbf A$'s eigenvalues $\{\lambda_k\}$ as
\begin{align*}
  b_{\mathrm c}(G) &=  \left(\sum_k \sinh (\lambda_k)\right) / \left(\sum_k
  e^{\lambda_k}\right), 
\end{align*}
This measure is always in the range $[0, 1/2]$, and attains the value of
zero if and only if $G$ is bipartite.  

\subsection{Comparison}
\begin{figure}
  \newcommand{\wTab}{0.17\textwidth}
  \newcommand{\wTabH}{0.173\textwidth}
  \newcommand{\wTabC}{0.187\textwidth}
  \newcommand{\wTabR}{0.1665\textwidth}
  \newcommand{\cora}{1.0cm}
  \newcommand{\cocol}{+0.14cm}
  \centering
  \begin{tabular}{c@{\hskip 0.1cm}r@{\hskip \cocol}c@{\hskip \cocol}c@{\hskip \cocol}c}
    \rotatebox{90}{{\hskip 0.7cm} $b_{\mathrm K}$}
    & \includegraphics[width=\wTab]{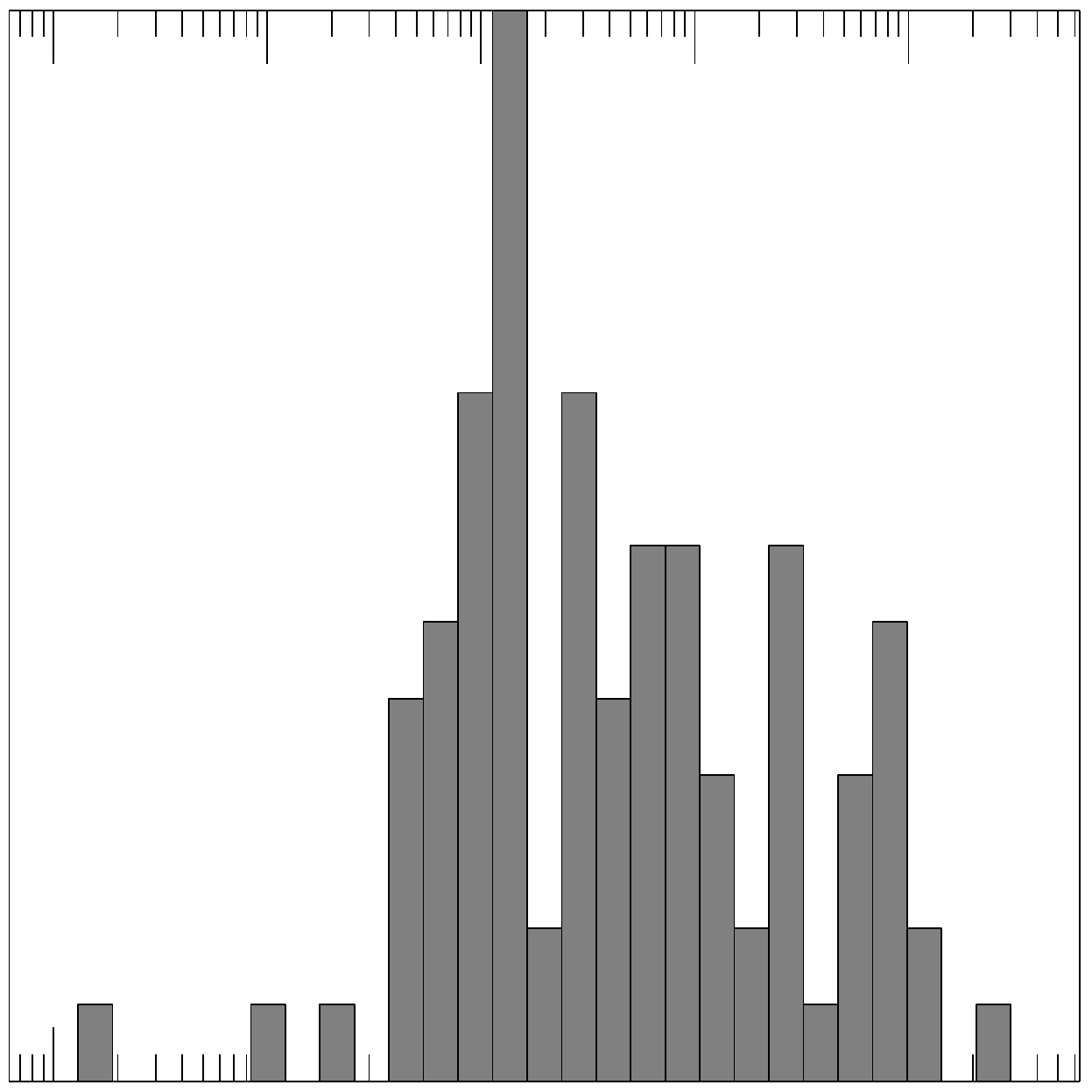}
    & \raisebox{\cora}{\input{tex-bipartivity/scattercorr.anticonflict.nonbip}}
    & \raisebox{\cora}{\input{tex-bipartivity/scattercorr.anticonflict.nonbipn}}
    & \raisebox{\cora}{\input{tex-bipartivity/scattercorr.anticonflict.oddcycles}}
    \\
    \rotatebox{90}{{\hskip 0.7cm} $b_{\mathrm A}$}
    & \includegraphics[height=\wTabH]{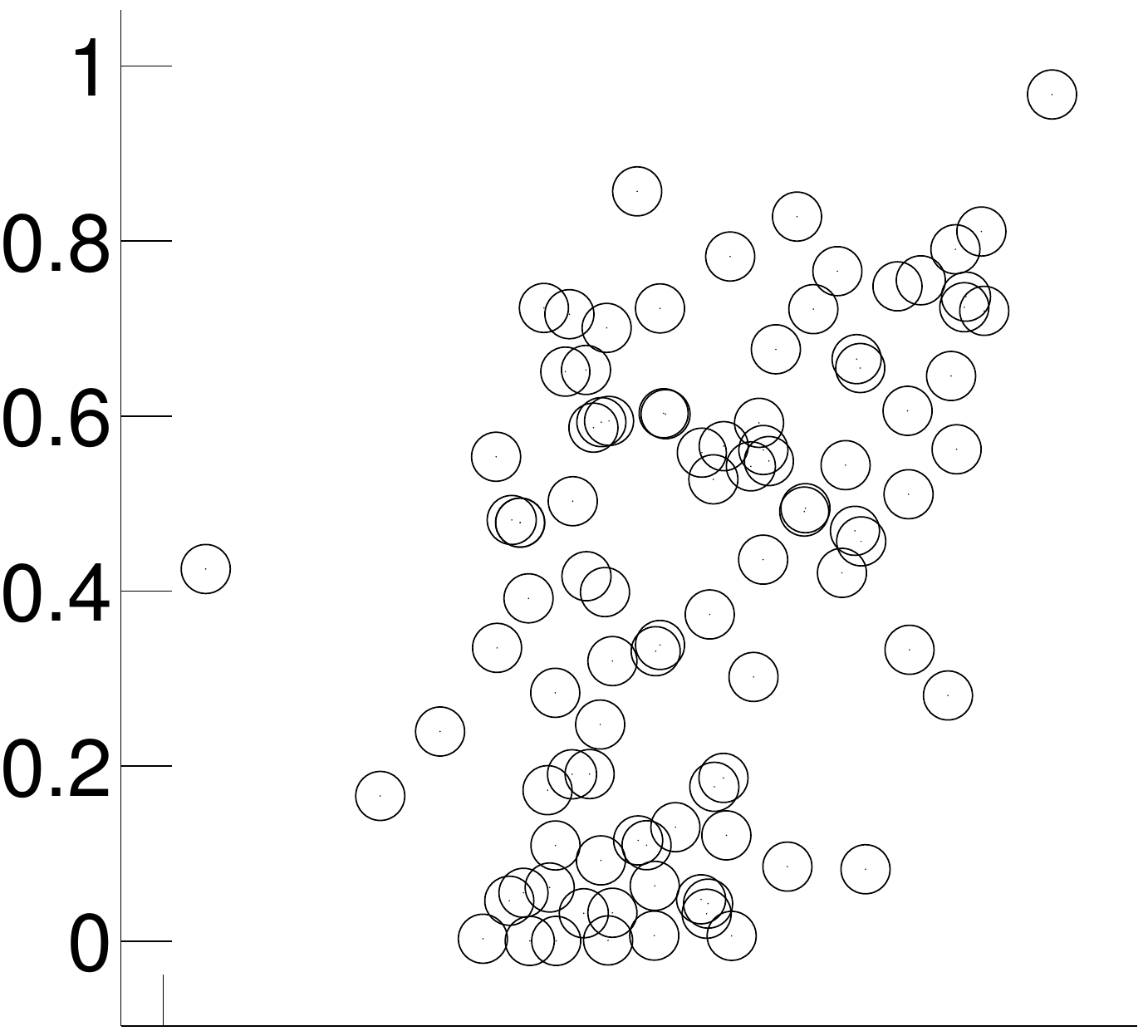}
    & \includegraphics[width=\wTab]{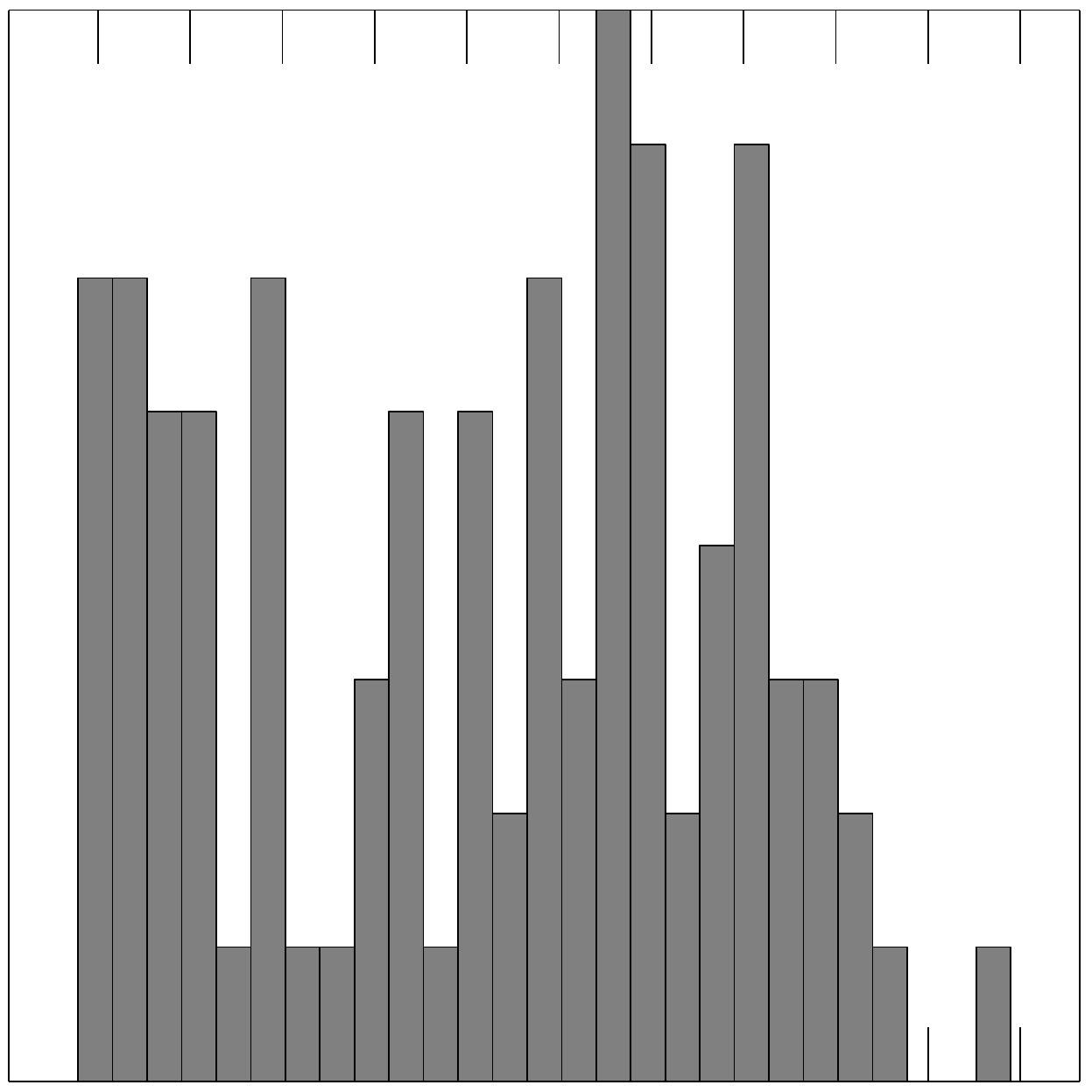} 
    & \raisebox{\cora}{\input{tex-bipartivity/scattercorr.nonbip.nonbipn}}
    & \raisebox{\cora}{\input{tex-bipartivity/scattercorr.nonbip.oddcycles}}
    \\
    \rotatebox{90}{{\hskip 0.8cm} $b_{\mathrm N}$}
    & \includegraphics[height=\wTabH]{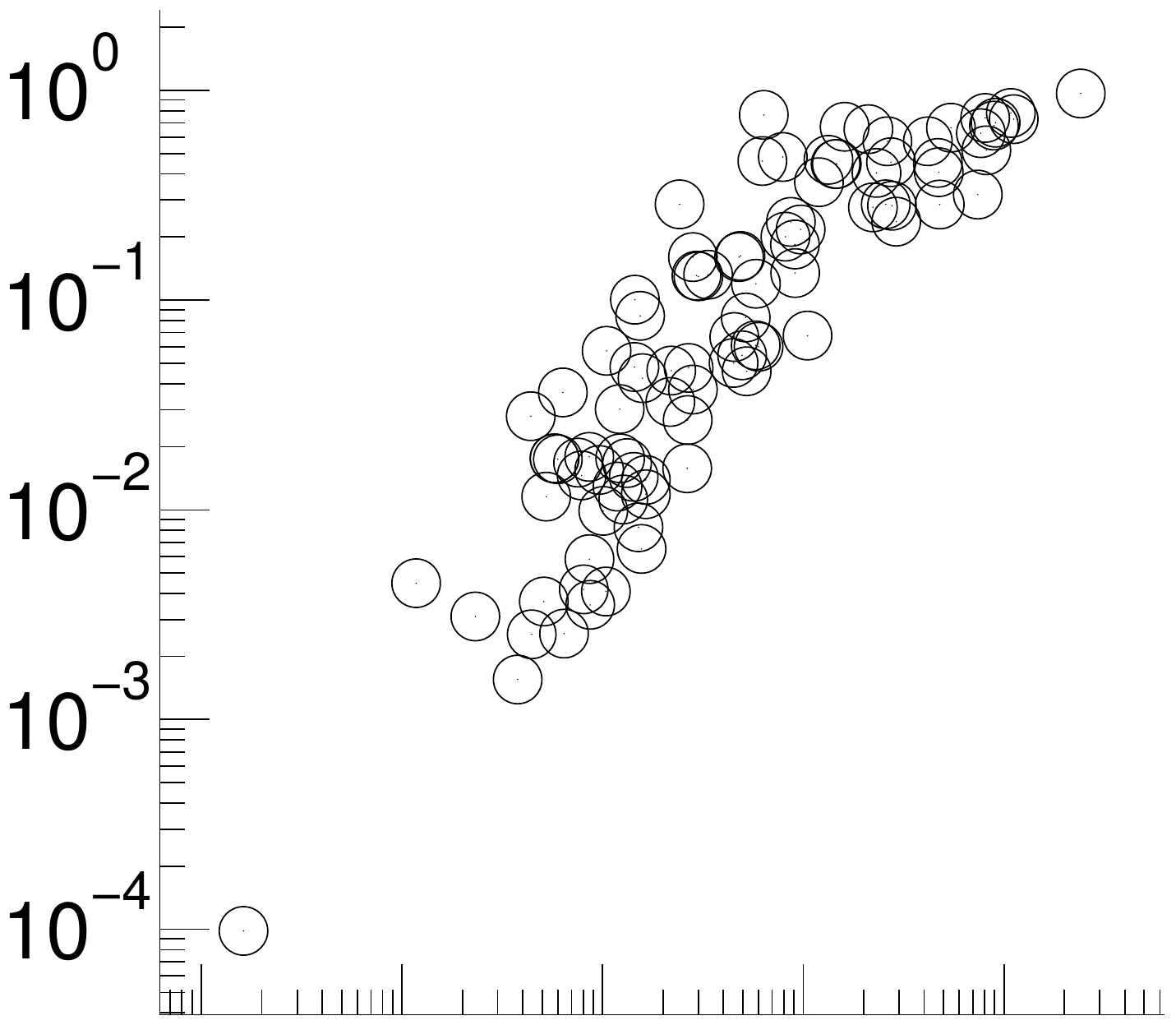}
    & \includegraphics[width=\wTab]{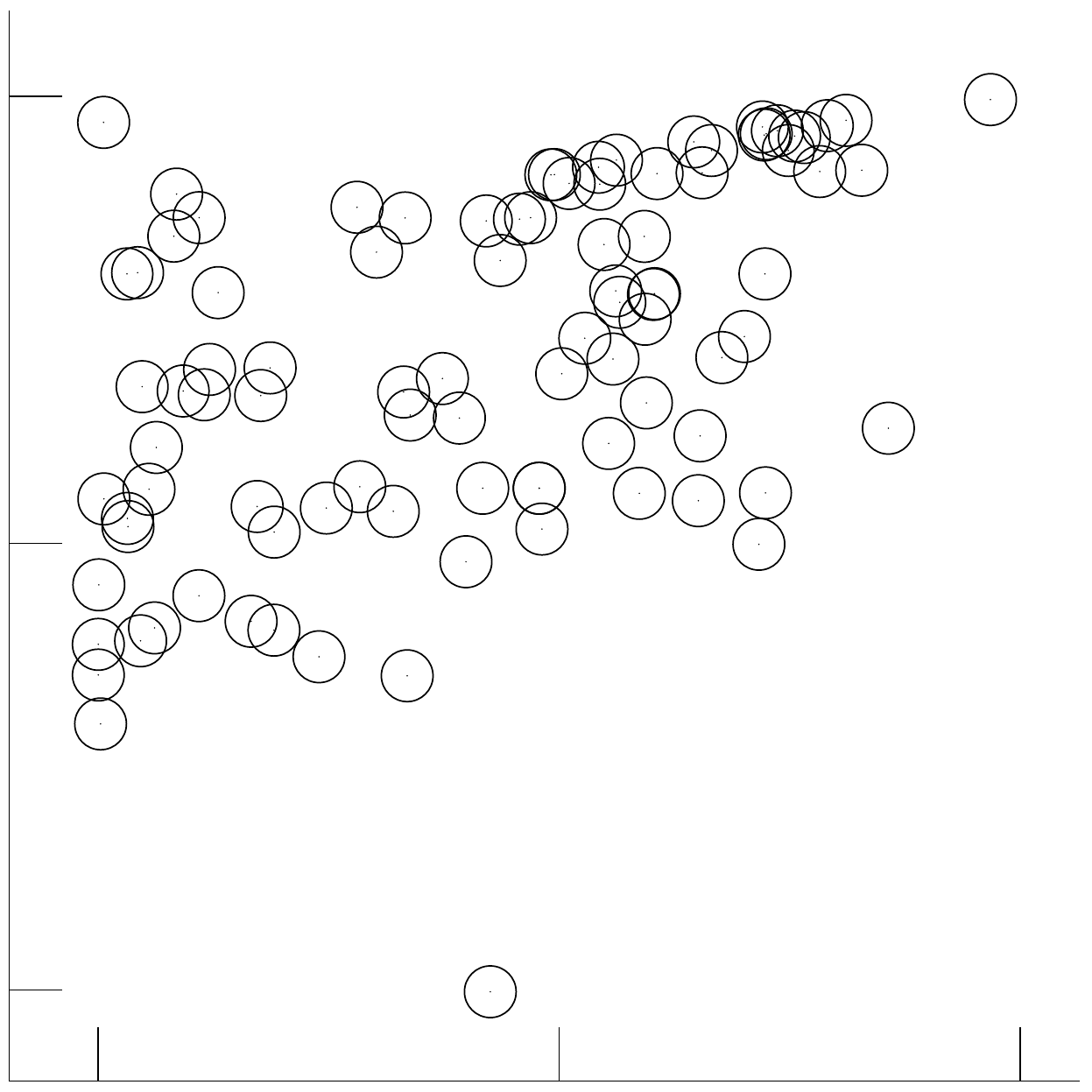}
    & \includegraphics[width=\wTabR]{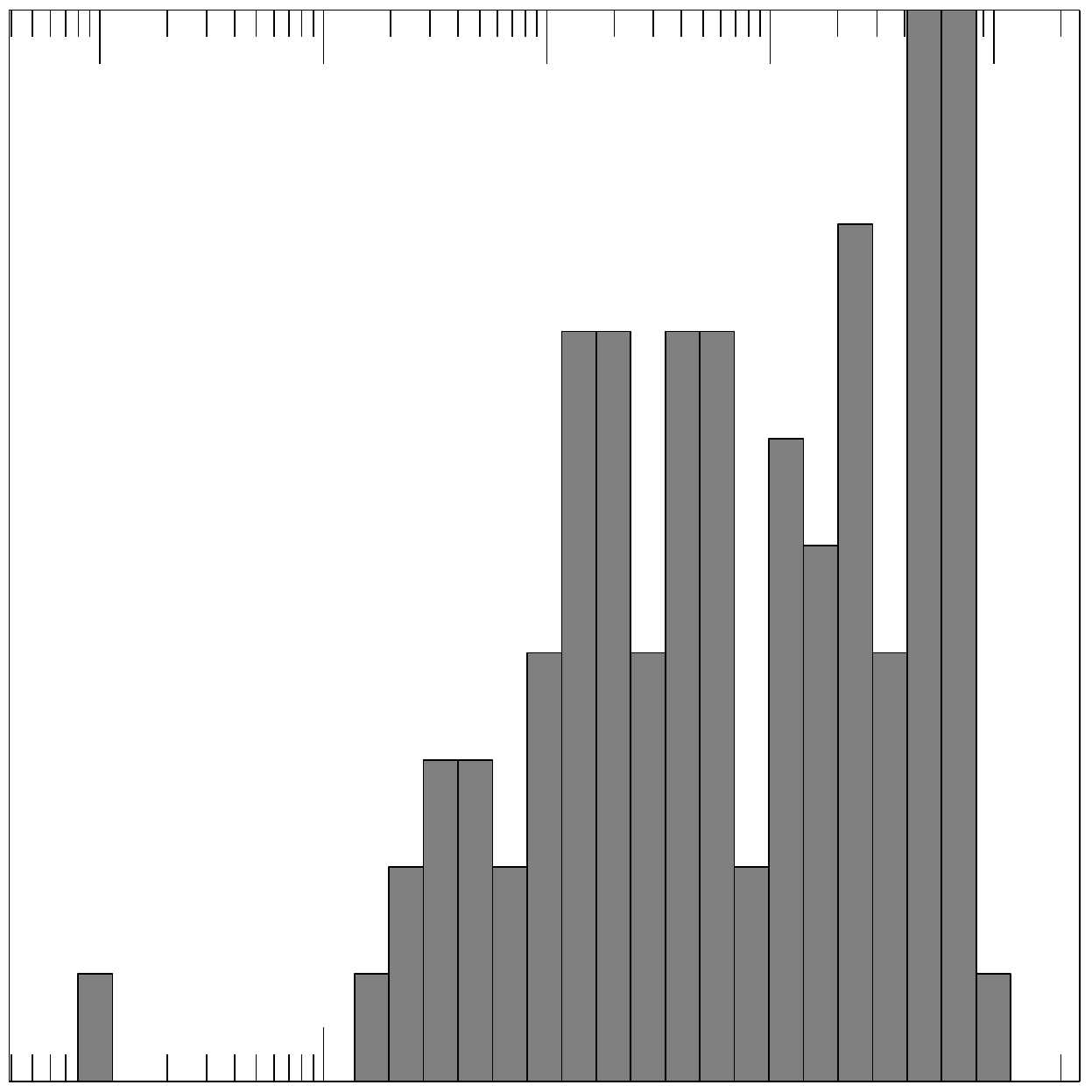}
    & \raisebox{\cora}{\input{tex-bipartivity/scattercorr.nonbipn.oddcycles}}
    \\
    \rotatebox{90}{{\hskip 0.9cm} $b_{\mathrm c}$}
    & \includegraphics[height=\wTabC]{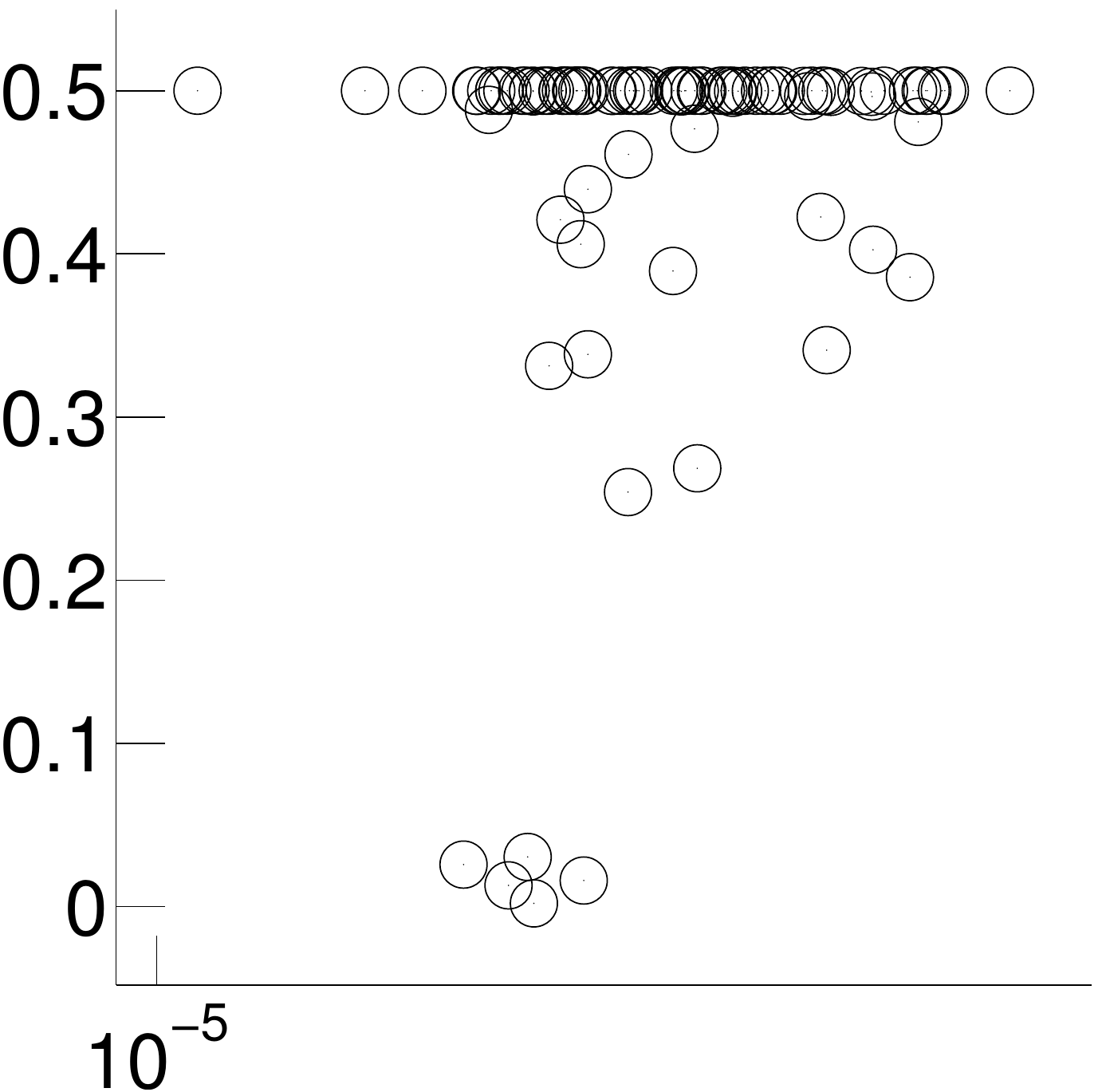} 
    & \includegraphics[height=\wTabC]{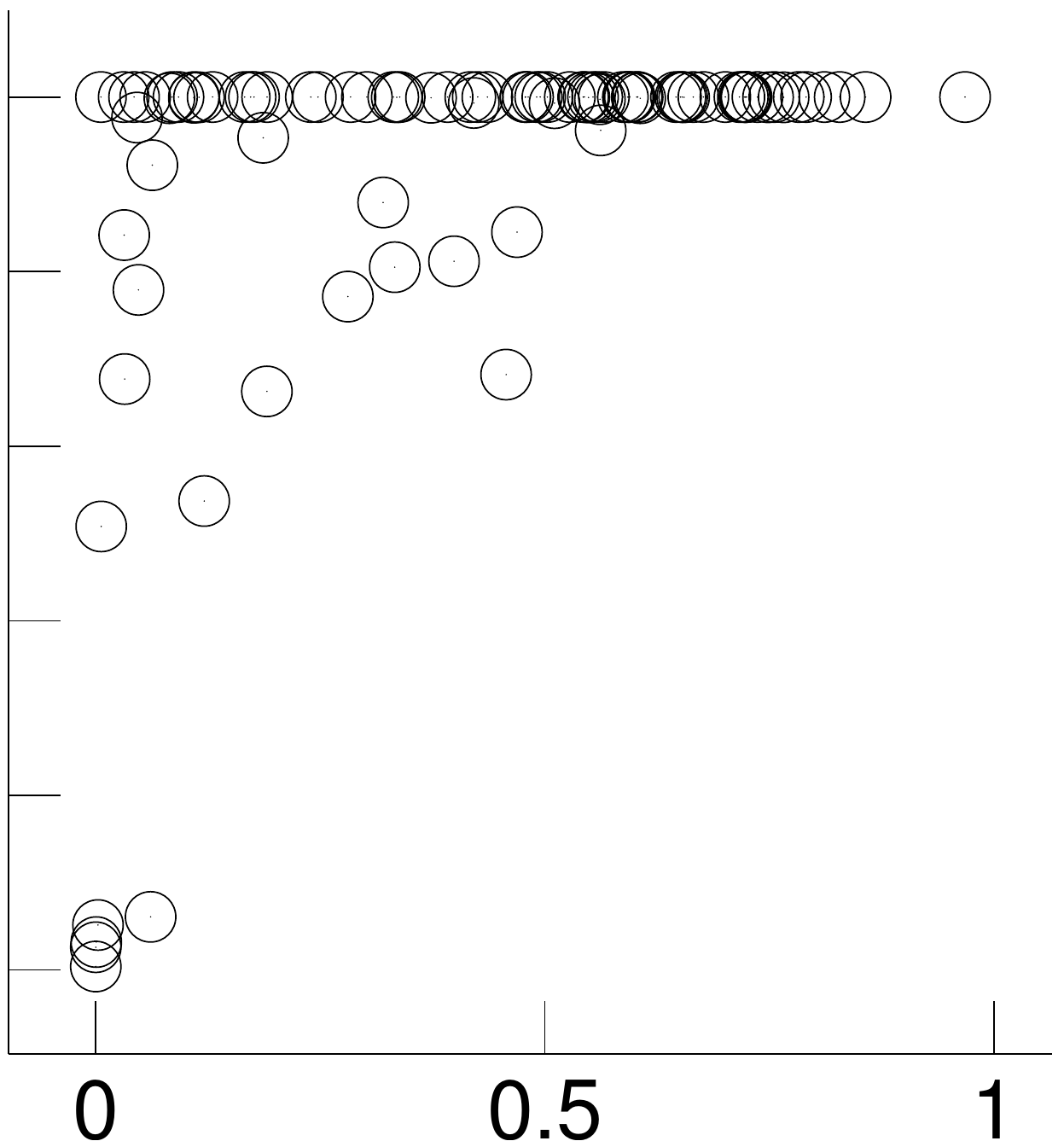} 
    & \includegraphics[height=\wTabC]{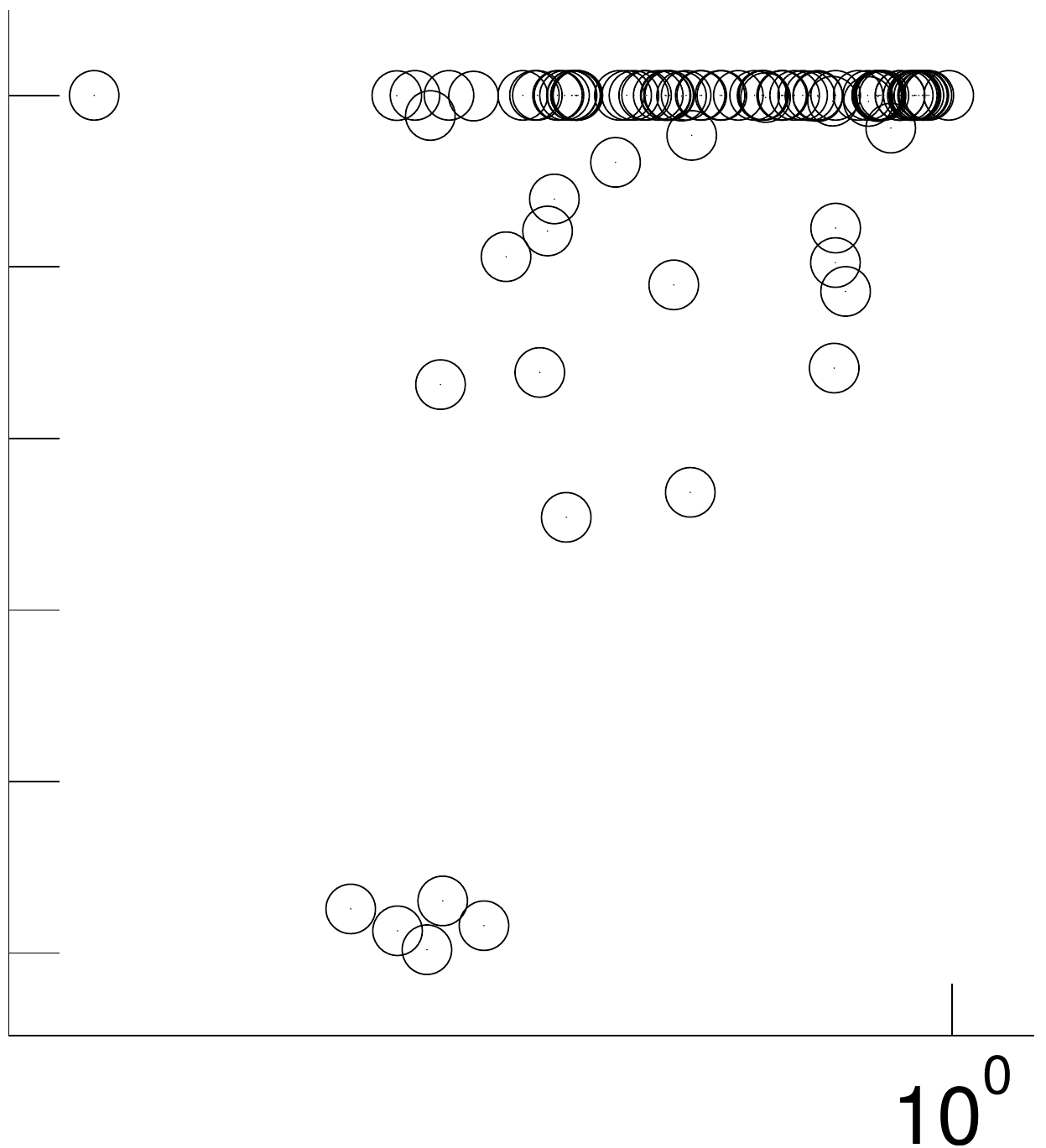} 
    & \includegraphics[width=\wTab]{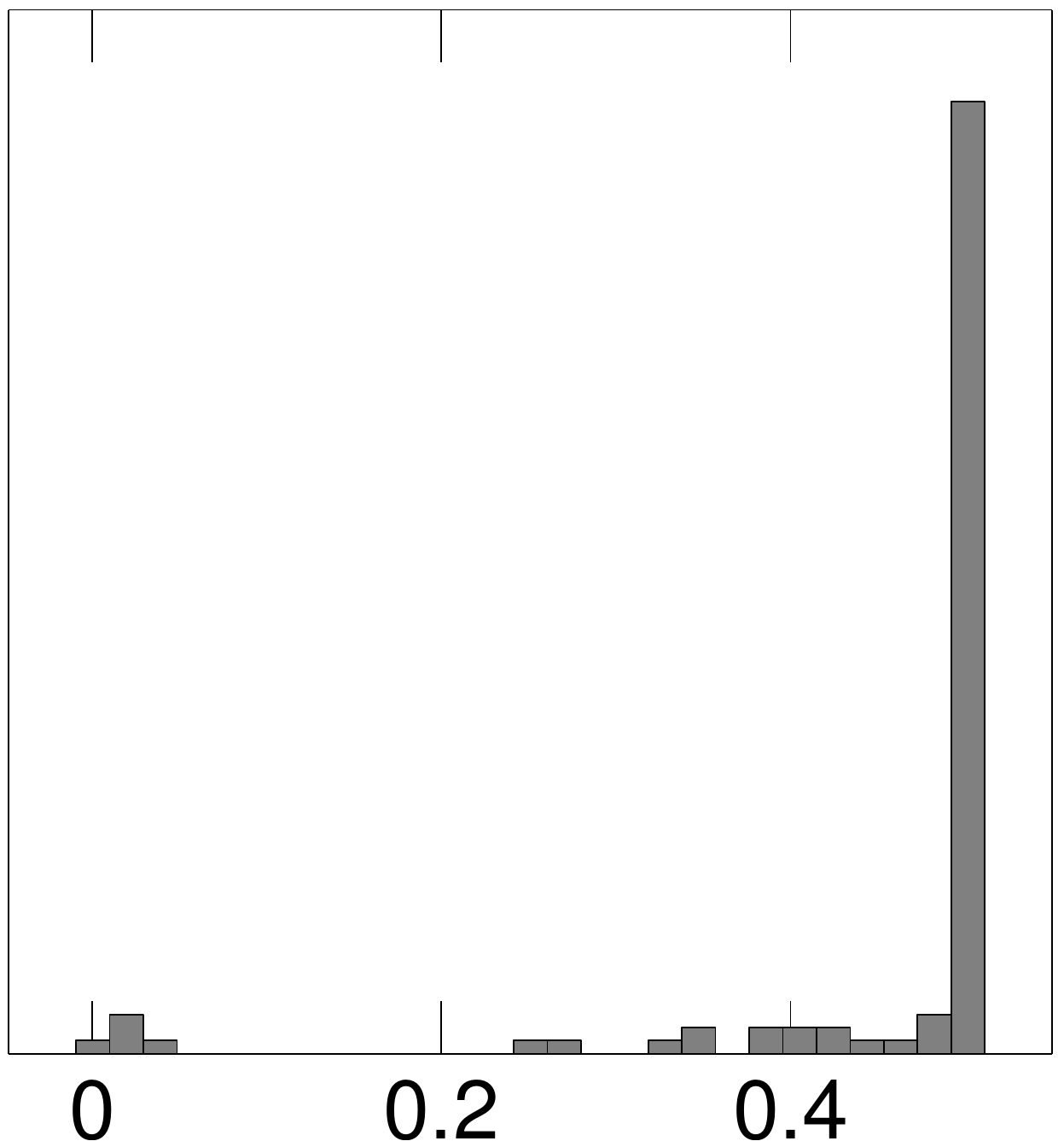} 
    \\[-0.05cm]
    & \multicolumn{1}{c}{$b_{\mathrm K}$} & $b_{\mathrm A}$ & $b_{\mathrm N}$ & $b_{\mathrm c}$
  \end{tabular}
  \caption{
    Quantitative comparison of the three measures of non-bipartivity.
    The diagonal subplots show histograms for each measure.  The plots
    under the diagonal show scatter plots for each pair of
    measures. Over the diagonal, the Pearson correlation coefficients
    $\rho$ are shown, with $p$-values lower than 0.1, 0.01 and 0.001
    shows as one, two and three stars.
    \label{fig:nonbip}
  }
\end{figure}

\begin{figure}
  \centering
  \includegraphics[width=0.6\textwidth]{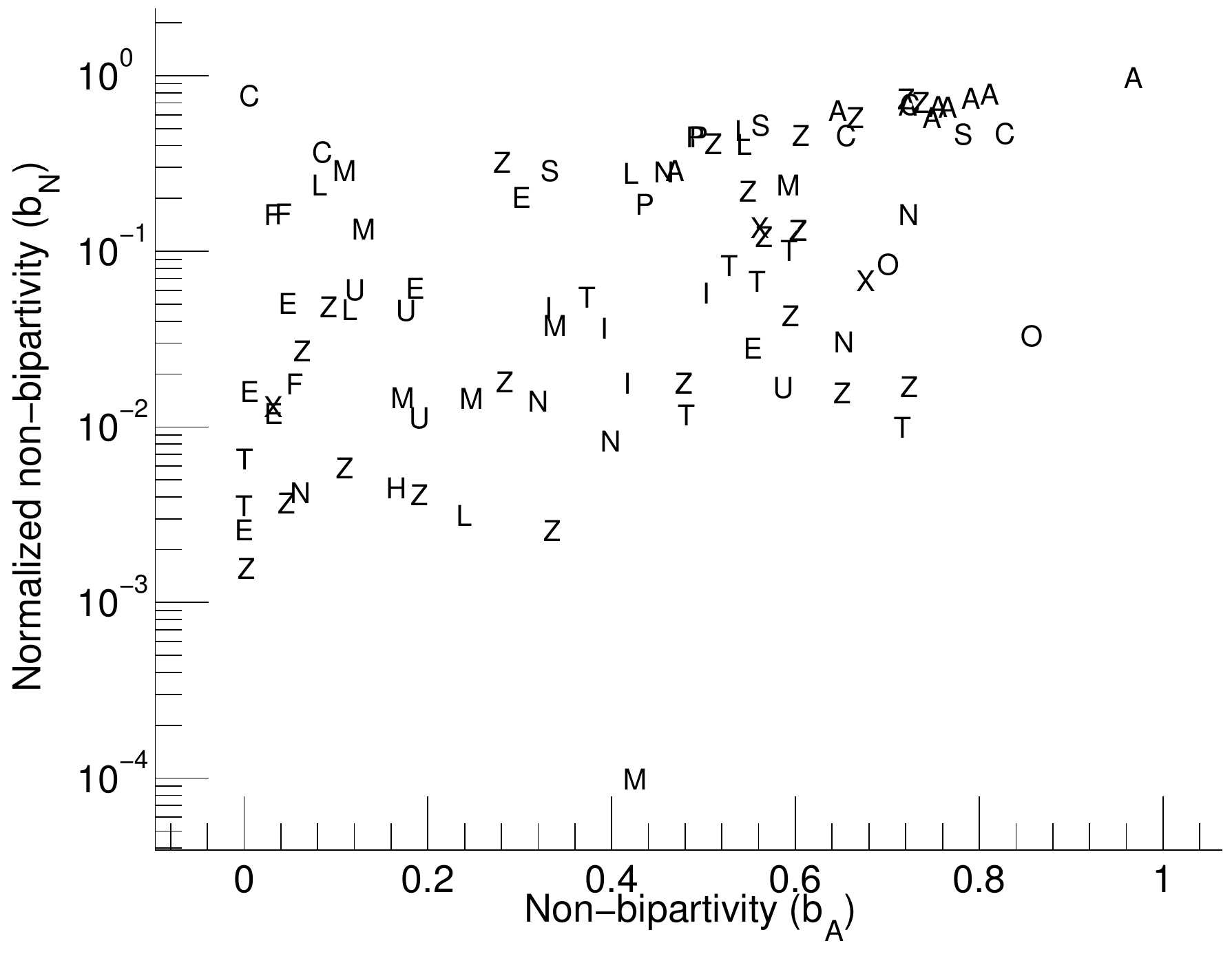}
  \raisebox{3cm}{
    \scalebox{0.7}{
    \begin{tabular}{ c l }
      \textsf{A} & Animal network \\
      \textsf{C} & Human contact network \\
      \textsf{E} & Metabolic network \\
      \textsf{F} & Software network \\
      \textsf{H} & Hyperlink network \\
      \textsf{I} & Citation network \\
      \textsf{L} & Lexical network \\
      \textsf{M} & Communication network \\
      \textsf{N} & Infrastructure network \\
      \textsf{O} & Coauthorship network \\
      \textsf{P} & Trophic network \\
      \textsf{S} & Social network \\
      \textsf{T} & Online contact network \\
      \textsf{U} & Computer network \\
      \textsf{X} & Miscellaneous network \\
      \textsf{Z} & Online social network 
    \end{tabular}
    }
  }
  \caption{
    \label{fig:nonbip.nonbipn}
    All analysed networks plotted by the non-bipartivity measures
    $b_{\mathrm A}$ and $b_{\mathrm N}$. 
    Each letter code stands for a single one-mode network. 
    The individual letters correspond to the type of network. 
  }
\end{figure}

A comparison of the three measures of non-bipartivity is given in
Table~\ref{tab:nonbip}, and a numerical comparison of them is shown in
Figure~\ref{fig:nonbip}. A detailed comparison plot of the two measures
$b_{\mathrm A}$ and $b_{\mathrm N}$ is shown in
Figure~\ref{fig:nonbip.nonbipn}.  The numerical comparison shows that
the values of $b_{\mathrm c}$ (the ratio of odd cycles) are in fact very
near to $0.5$ for almost all networks, making the measure useless for
characterizing large networks.

Figure~\ref{fig:nonbip.nonbipn} shows that bipartivity does not
correlate with the type of network.  While the most bipartite networks
according to both measures are reference networks (the Notre Dame web
(\href{http://konect.uni-koblenz.de/networks/web-NotreDame}{\textsf{ND}})
and Baidu related pages
(\href{http://konect.uni-koblenz.de/networks/zhishi-baidu-relatedpages}{\textsf{BAr}})),
other reference networks have very different non-bipartivity values.
The same is also true for other network categories.

The runtimes needed to compute each of the four measures of
non-bipartivity can be derived as follows.  The fastest one to compute
is the one based on the spectrum of the adjacency matrix $b_{\mathrm
  A}$, since we need only compute the maximum and minimum eigenvalues of
$\mathbf A$, which can be done efficiently even for very large graphs.
For $b_{\mathrm N}$, we must compute the minimum eigenvalue of the
normalized adjacency matrix $\mathbf N$.  Since $\mathbf N$'s
eigenvalues are much less separated than $\mathbf A$'s, a power
iteration on $\mathbf N$ takes longer than on $\mathbf A$.  Computing
the minimum eigenvalue of $\mathbf K$ is even slower, and needs more
memory, as a sparse LU decomposition of $\mathbf K$ must be computed.
Computing the measure $b_{\mathrm c}$ is high too, as it needs all
eigenvalues of $\mathbf A$.

From the histograms shown in Figure~\ref{fig:nonbip}, we can conclude
that only $b_{\mathrm A}$ has an almost-uniform distribution, and thus
we suppose it to be more informative.  Since it also is the fastest of
the measures to compute, we recommend to compute $b_{\mathrm A}$, i.e.,
$1 - |\lambda_{\min}[\mathbf A] / \lambda_{\max}[\mathbf A]|$ as a
measure of bipartivity for one-mode networks.

\begin{table}
  \caption{
    \label{tab:nonbip}
    Qualitative comparison of the four measures of non-bipartivity. 
  }
  \centering
  \begin{tabular}{lp{0.4\textwidth}ll}
    \toprule
    \textbf{Measure} & \textbf{Measured feature} & \textbf{Range} 
    & \textbf{Runtime} \\ 
    \midrule

    Frustration ($b_{\mathrm K}$) & 
    Relaxed proportion of edges that must be removed to make
    the graph bipartite & 
    $[0, 1/2]$ & 
    Very high \\ 

    Spectrum of $\mathbf A$ ($b_{\mathrm A}$) & 
    Distance of smallest eigenvalue of the adjacency matrix to the
    opposite of the largest eigenvalue &
    $[0,1]$ &
    Low \\

    Spectrum of $\mathbf N$ ($b_{\mathrm N}$) & 
    Distance of smallest eigenvalue of the normalized adjacency matrix to the
    opposite of the largest eigenvalue &
    $[0,1]$ &
    Middle \\

    Odd cycles ($b_{\mathrm c}$) & 
    Proportion of odd cycles to total cycles, weighted by
    inverse factorial of cycle length & 
    $[0, 1/2]$ & 
    High \\

    \bottomrule
  \end{tabular}
\end{table}

\section{Clustering Analysis}
\label{sec:clustering}
The term \emph{clustering} refers to the observation that real-world
networks, for instance social networks, tend to contain clusters, i.e.,
groups of nodes with a large number of edges among them, and few nodes
connected to the outside of the group.  A major problem class in machine
learning, and indeed in unsupervised learning is the task of
\emph{clustering}. Given a set of data points and their attributes, the
clustering problem asks to find subsets of these data points that are
similar, in a way depending on the attributes.  Examples of applications
that can be reduced to clustering include community detection in social
networks and the unsupervised determination of topics in a scientific
collaboration graph.

The most general formulation of the clustering problem uses pairwise
distances between the points. As an example, the $k$-means algorithm
finds a partition of a set of points such that each point is nearest to
the centroid of its assigned cluster.

If the points to be clustered are vertices in a graph, and the edges
connecting the vertices are the only information available, then one
speaks of \emph{graph clustering}~\cite{b304}.  Of the many graph
clustering methods in existence, we will review a spectral one that is
both very general and which performs reasonably well in practice:
ratio-cut spectral clustering. We will also restrict our treatment to
the case of 2-clustering, i.e., finding a partition of the vertex set of
a graph into two parts, such that each part contains as many edges as
possible, and only few edges connect the two parts.

\begin{figure}
  \centering
  \includegraphics[width=\wTwo]{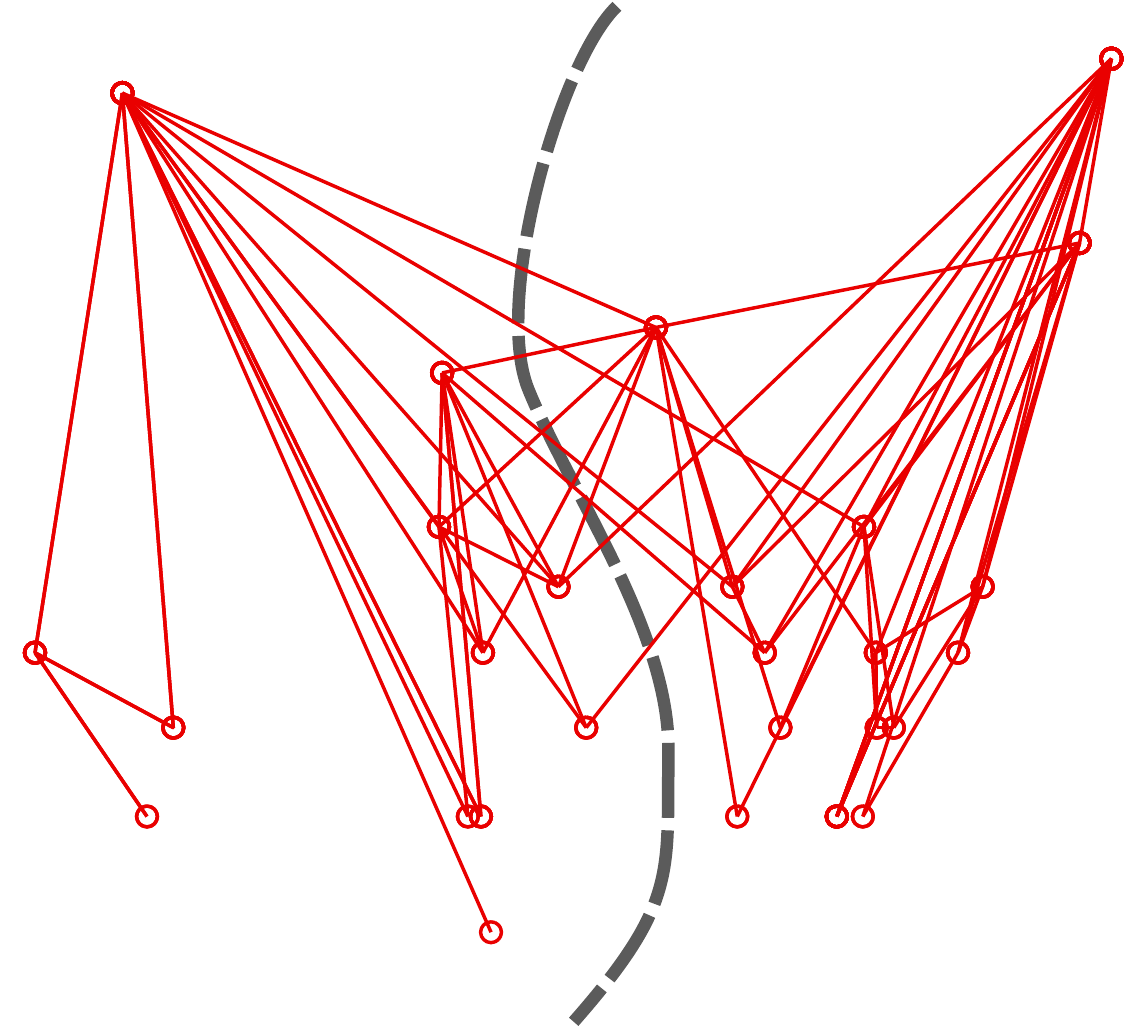}
  \caption{
    The optimal 2-clustering of Zachary's unipartite karate club network,
    using the ratio cut criterion. 
  }
  \label{fig:clustering-uni}
\end{figure}

Let $G=(V,E)$ be a connected unipartite graph. Given any partition $V=X
\cup Y$, the cut of the pair $(X, Y)$ is defined as the number of edges
connecting $X$ and~$Y$:
\begin{align*}
  \mathrm{cut}(X,Y) &= |\left\{ u, v \in V \mid u \in X, v \in Y \right\}|
\end{align*}
A good 2-clustering will thus give a low value of $\mathrm{cut}(X,
Y)$. Using the cut as a minimization objective does not work however,
since the result would be very skewed to partitions with one very small
and one very large cluster. Thus, an suitable objective function for
minimization is given by the ratio cut~\cite{b304}
\begin{align*}
  \mathrm{rcut}(X,Y) &= \left(\frac 1 {|X|} + \frac 1 {|Y|}\right) \mathrm{cut}(X,
  Y). 
\end{align*}
The ratio cut can be minimized by expressing is in terms of the vector
$\mathbf x \in \mathbb R^{|V|}$ defined by
\begin{align}
  \mathbf x_u &= + \sqrt{|Y| / |X|} \text{ when } u \in X \label{eq:cut} \\
  \mathbf x_u &= - \sqrt{|X| / |Y|} \text{ when } u \in Y. \nonumber
\end{align}
Using the Laplacian matrix $\mathbf L\in \mathbb R^{|V|\times |V|}$
defined by $\mathbf L_{uu}=d(u)$, $\mathbf L_{uv}=-1$ when $u$ and $v$
are connected and $\mathbf L_{uv}=0$ otherwise, we can check that
\begin{align*}
  \mathbf x \mathbf L \mathbf x^{\mathrm T} &= 2|V| \mathrm{rcut}(X, Y)
\end{align*}
and that $\sum_u \mathbf x_u=0$, i.e., $\mathbf x$ is orthogonal to the
constant vector. Denoting by $\mathcal C$ the vectors of the form given
in Equation~\eqref{eq:cut}, the clustering problem becomes
\begin{align*}
  \min_{\mathbf x \in \mathbb R^{|V|}} & \quad \mathbf x \mathbf L \mathbf
  x^{\mathrm T} \\
  \mathrm{s.t.} & \quad \mathbf x \perp \mathbf 1, \mathbf x \in \mathcal C
\end{align*}
This can be relaxed by removing the constraint $\mathbf x \in \mathcal
C$, giving as a solution the eigenvector of $\mathbf L$ having the
smallest nonzero eigenvalue.  The optimal $\mathbf x$ can then be used
for 2-clustering, by defining the two clusters to comprise the nodes
with positive and negative values in $\mathbf x$, respectively.
Equivalently, the optimal $\mathbf x$ can be found by computing $\mathbf
x = \mathbf D^{-1/2}\mathbf y$, where $\mathbf D$ is the diagonal degree
matrix defined by $\mathbf D_{uu}=d(u)$, and $\mathbf y$ is the
eigenvector corresponding to the second largest absolute eigenvalue of
the matrix $\mathbf D_1^{-1/2} \mathbf A \mathbf D_2^{-1/2}$. Note that
the largest eigenvalue of that matrix is one, and the corresponding
eigenvector has entries inversely proportional to the vertex degrees.
Here, $\mathbf D_1$ and $\mathbf D_2$ refer to the diagonal degree
matrices computed from $\mathbf D$ by keeping all the left and right
vertices, respectively.  As an example, the resulting 2-clustering of
Zachary's unipartite karate club network is shown in
Figure~\ref{fig:clustering-uni}.

\begin{figure}
  \centering
  \includegraphics[width=\wOnePointFive]{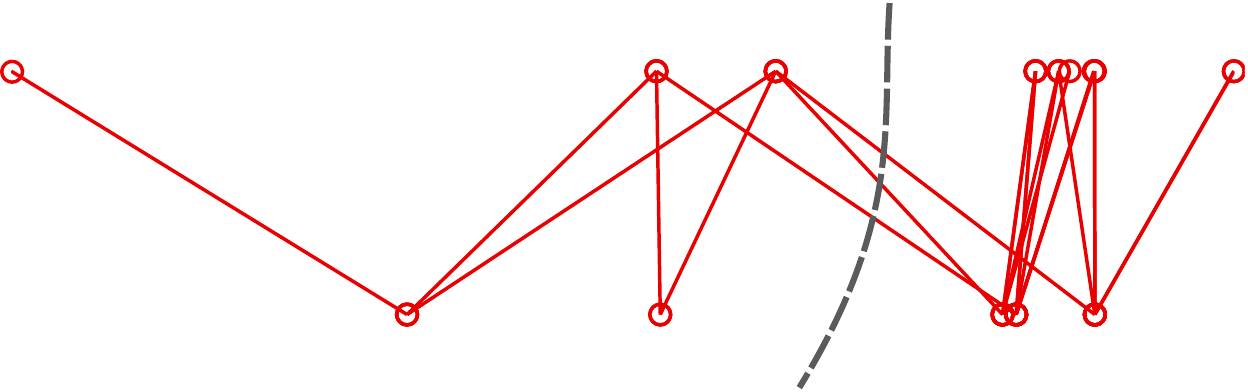}
  \caption{
    The optimal 2-clustering of the bipartite artist--genre network,
    using the ratio cut criterion. 
  }
  \label{fig:clustering-bip}
\end{figure}

In the case of a bipartite graph $G(V_1,V_2,E)$, the problem of
clustering is usually called biclustering or co-clustering~\cite{b309}.
These names refer to the fact that a good clustering of a bipartite
graph must cluster both vertex sets in a way that is compatible with
each other.  In other words, the graph biclustering problem consists in
finding the partitions
\begin{align*}
  V_1 &= X_1 \cup Y_1, \\
  V_2 &= X_2 \cup Y_2.
\end{align*}
In the case of the ratio cut, the objective is thus to minimize
$\mathrm{rcut}(X_1 \cup X_2, Y_1 \cup Y_2)$.  This can be realized by
writing the network's normalized adjacency matrix as
\begin{align*}
  \mathbf D^{-1/2} \mathbf A \mathbf D^{-1/2} &= 
  \left[ \begin{array}{cc} 
      \mathbf 0 &  \mathbf D_1^{-1/2} \mathbf B \mathbf D_2^{-1/2} \\
      \mathbf D_2^{-1/2} \mathbf B^{\mathrm T} \mathbf D_1^{-1/2} & \mathbf 0 \\
    \end{array} \right],
\end{align*}
in which $\mathbf B$ is the biadjacency matrix of the network.  It
follows that a 2-clustering of a bipartite graph based on ratio cuts can
be computed by finding the left and right singular vectors of the matrix
$\mathbf D_1^{-1/2} \mathbf B \mathbf D_2^{-1/2}$ corresponding to the
second-largest eigenvalue.  As an example, the resulting 2-clustering of
the bipartite artist--genre network is shown in
Figure~\ref{fig:clustering-bip}.

\subsection{Other Approaches}
Another way to generalize the problem of graph clustering is by
considering the concept of a clique.  A clique in a graph is a set of
nodes all of which are connected to each other. In other words, a clique
is a complete graph $K_n$ that appears as a subgraph of another graph.
In the bipartite case, the equivalent of a clique is a complete
bipartite graph (or \emph{biclique}) $K_{m,n}$ that appears as a
subgraph. Now, a problem related to graph clustering is the problem of
finding large cliques in a unipartite graph. Generalizing this to
bipartite graphs results in the problem of finding large
bicliques~\cite{b4}.

\section{Visualization}
\label{sec:visualization}
Visualization is an important part of many exploratory sciences.  Since
so many different types of data can be modeled as networks, it is not
surprising that network visualization constitutes a large topic.  In
fact, there are many different graph drawing algorithms, designed for
many different types of networks, many of which can also be applied to
bipartite networks. In this section, we describe one specific graph
drawing algorithm that ties directly to our previous exposition about
algebraic graph theory: spectral graph drawing, which consists in using
the eigenvectors of characteristic graph matrices to draw a network.  We
note that other graph drawing algorithms exists, and also other settings
in which they can be applied. However, we will restrict ourselves to the
case of two-dimensional graph drawing, i.e., the embedding of the nodes
of graph into a two-dimensional surface.

Let us begin by formalizing the graph drawing problem.  Let $G=(V,E)$ be
a graph.  In order to draw $G$, we need to determine the coordinates of
each node $u \in V$ in the drawing.  The edges can then be drawn as
lines from one node to another.  For simplicity, we will restrict
ourselves to graph drawing methods in which edges are drawn as straight
lines.  A graph drawing is by nature two-dimensional, and thus we must
find a mapping of the nodes into the plane $\mathbb R^2$.  We will
represent this mapping by a matrix $\mathbf X \in \mathbb R^{|V|\times
  2}$, in which each row $\mathbf X_{u:}$ represents the coordinates of
node $u$.

A chosen graph drawing can be required to fulfill many aesthetic
requirements, for instance that lines representing edges are not
overlapping, that the drawing has constant density of nodes over the
whole drawing, that cliques are apparent in the drawing, that nodes are
not drawn too near to each other, etc. There are many such requirements,
partially in conflict with each other.  Any choice of such requirements
(or rather, weighting of their importances), leads to a different graph
drawing algorithm. The requirement chosen as an example in this section
will be that of proximity: Nodes should be drawn near to their
neighbors. We chose this requirement because it reflects the importance
of the network structure over other aesthetic criteria, and because it
leads to a closed-form solution with can be exploited in practice, in
particular for very large networks.

\begin{figure}
  \centering
  \subfigure[Crossing-free drawing]{
    \includegraphics[width=\wTwo]{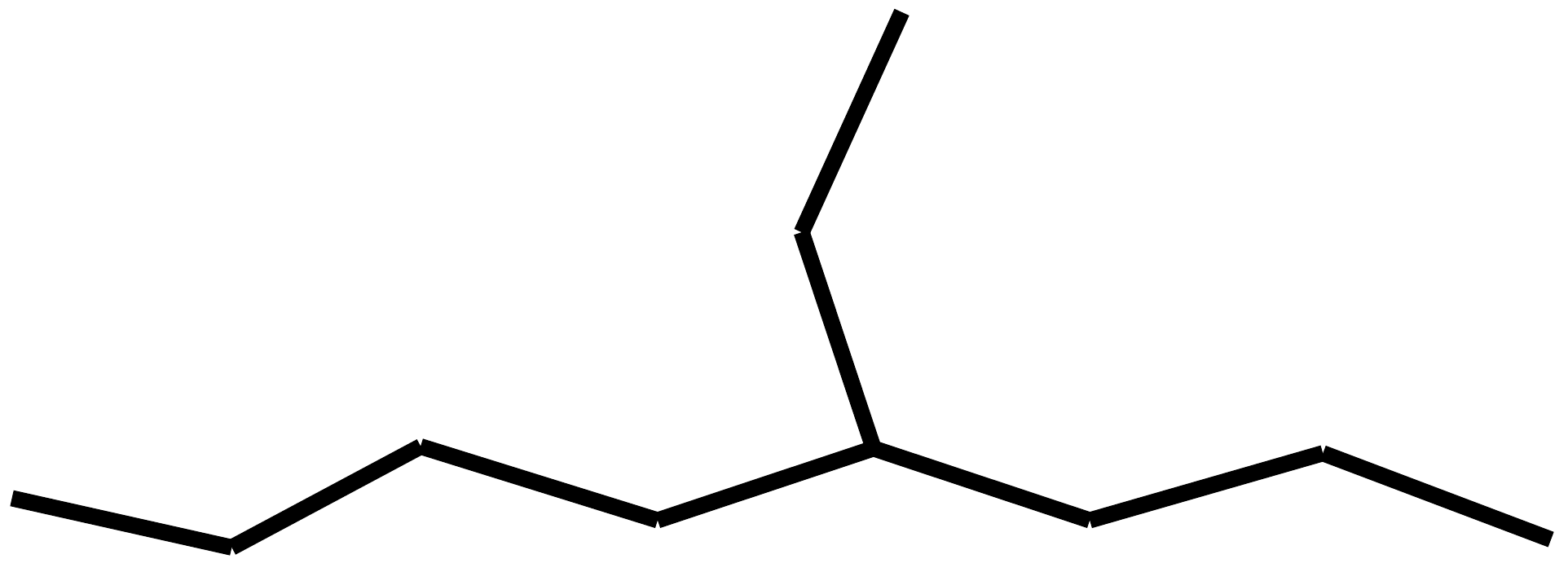}
    \label{fig:tree:no-overlap}
  }
  \subfigure[Highlighting the bipartite structure]{
    \includegraphics[width=\wTwo]{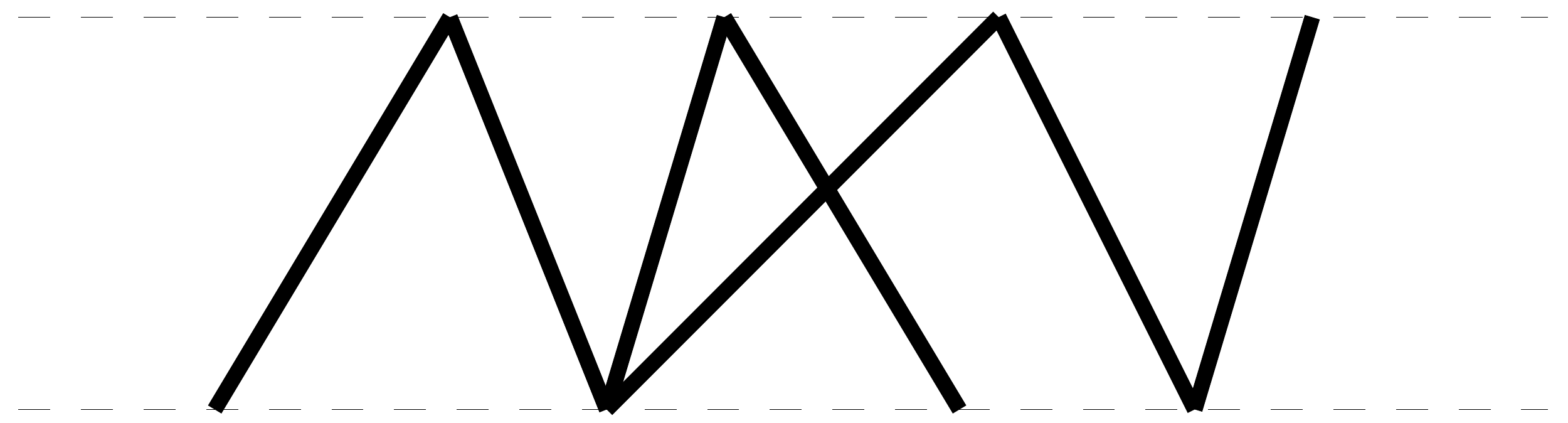}
    \label{fig:tree:bip}
  }
  \caption{
    An example of graph drawing that shows the conflict of aesthetic
    requirements: Drawing a tree (a connected cycle-free graph) can be
    done \subref{fig:tree:no-overlap}~without crossing edges but not
    showing the bipartite structure, or \subref{fig:tree:bip}~showing
    the bipartite structure but including crossing edges.  Note that it
    is impossible to fulfill both requirements at the same time for this
    graph, and therefore any bipartite graph drawing algorithm must
    choose which of these requirements will not be fulfilled in all
    cases.
  }
  \label{fig:tree}
\end{figure}

A second requirement which we will consider dictates that if a network
has special structure, that structure should be reflected in the graph
drawing. In the case of bipartite networks, this can for instance be
achieved by placing all nodes on two parallel lines in the drawing,
reflecting the two disjoint sets of nodes~\cite{b728}.  An illustration
of such a drawing is shown in Figure~\ref{fig:tree}, along with the
example of a tree (a connected cycle-free graph) which is bipartite, but
cannot be drawn in a nonintersecting manner when all nodes are placed on
two lines. This example shows that aesthetic requirements can be in
conflict with each other.

In the rest of section, we first define the Laplacian-based graph
drawing algorithm for unipartite graphs, and then show how it can be
extended to the case of bipartite graphs, with and without the
requirement to place the nodes on two parallel lines.

\subsection{Spectral Drawing of Unipartite Graphs}
We now describe the general method for generating an embedding of the
nodes of a graph into the plane using the Laplacian matrix.  Given a
connected graph $G=(V,E)$, its adjacency matrix $(\mathbf A_{uv})$ gives
the positive edge weights when the vertices $u$ and $v$ are connected,
and is zero otherwise.  We now want to find a two-dimensional drawing of
the graph in which each vertex is drawn near to its neighbors.  This
requirement gives rise to the following vertex equation, which states
that every vertex is placed at the mean of its neighbors' coordinates,
weighted by the weight of the connecting edges.  For each node $u$, let
$\mathbf X_{u:} \in \mathbb R^2$ be its coordinates in the drawing, then
\begin{align*}
  \mathbf X_{u:} &= \left(\sum_{v \sim u} \mathbf A_{uv}\right)^{-1} \sum_{v \sim u}
  \mathbf A_{uv} \mathbf X_{v:}. 
\end{align*}
Rearranging and aggregating the equation for all $u$ we arrive at
\begin{align*}
  \mathbf D \mathbf X &= \mathbf A \mathbf X \\
  \mathbf L \mathbf X &= \mathbf 0 
\end{align*}
In other words, the columns of $\mathbf X$ should belong to the null
space of the Laplacian matrix $\mathbf L$, which leads to the degenerate
solution of $\mathbf X$ containing constant vectors, as the constant
vector is an eigenvector of $\mathbf L$ with eigenvalue $0$.  To exclude
that solution, we require additionally that the column vectors of
$\mathbf X$ are orthogonal to the constant vector and to each other,
leading to the columns of $\mathbf X$ being the eigenvectors associated
with the two smallest eigenvalues of $\mathbf L$ different from zero.
This solution results in a well-known satisfactory embedding of
positively weighted graphs:
\begin{align*}
  \mathbf X_{: 1} &= \mathbf x_2[\mathbf L] \\
  \mathbf X_{: 2} &= \mathbf x_3[\mathbf L]
\end{align*}
where $\mathbf x_k[\mathbf L]$ denotes the eigenvector corresponding to
the $k$'s smallest eigenvalue of $\mathbf L$.  Such an embedding is
related to the resistance distance (or commute time distance) between
nodes of the graph~\cite{b287}.

\begin{figure}
  \centering
  \subfigure[Zachary's karate club]{
    \includegraphics[width=\wTwo]{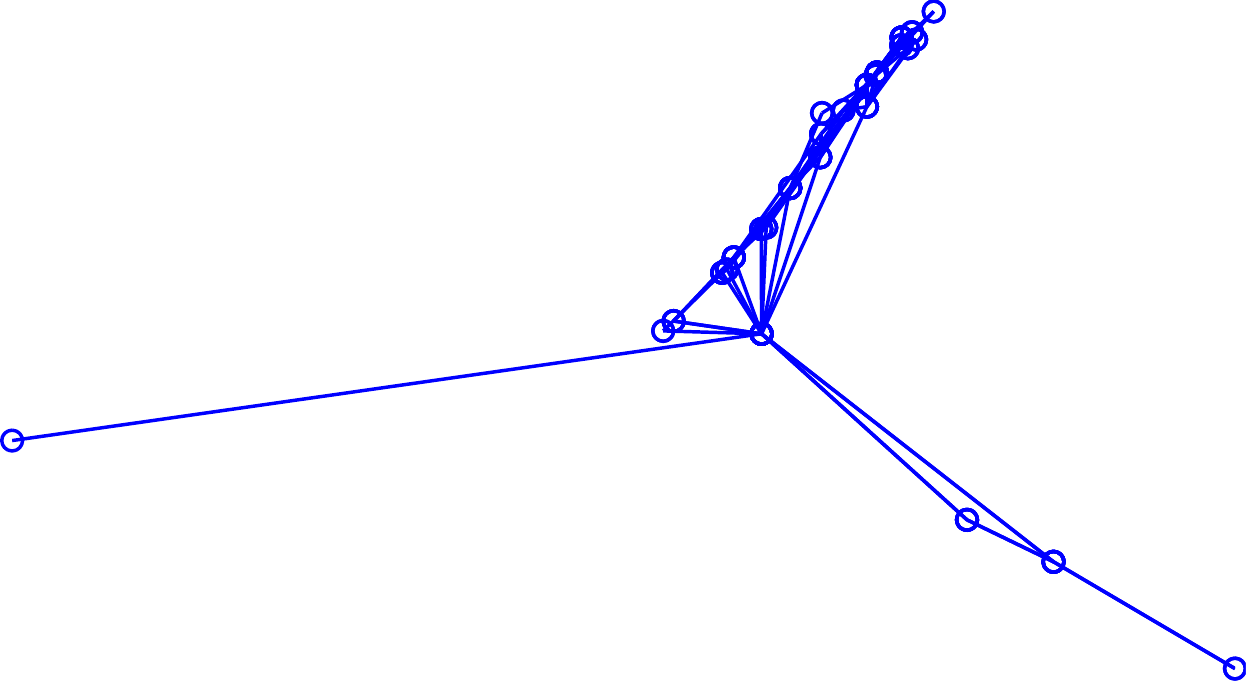}
    \label{fig:drawing-uni:zachary}
  }
  \subfigure[U.S. power grid]{
    \includegraphics[width=\wTwo]{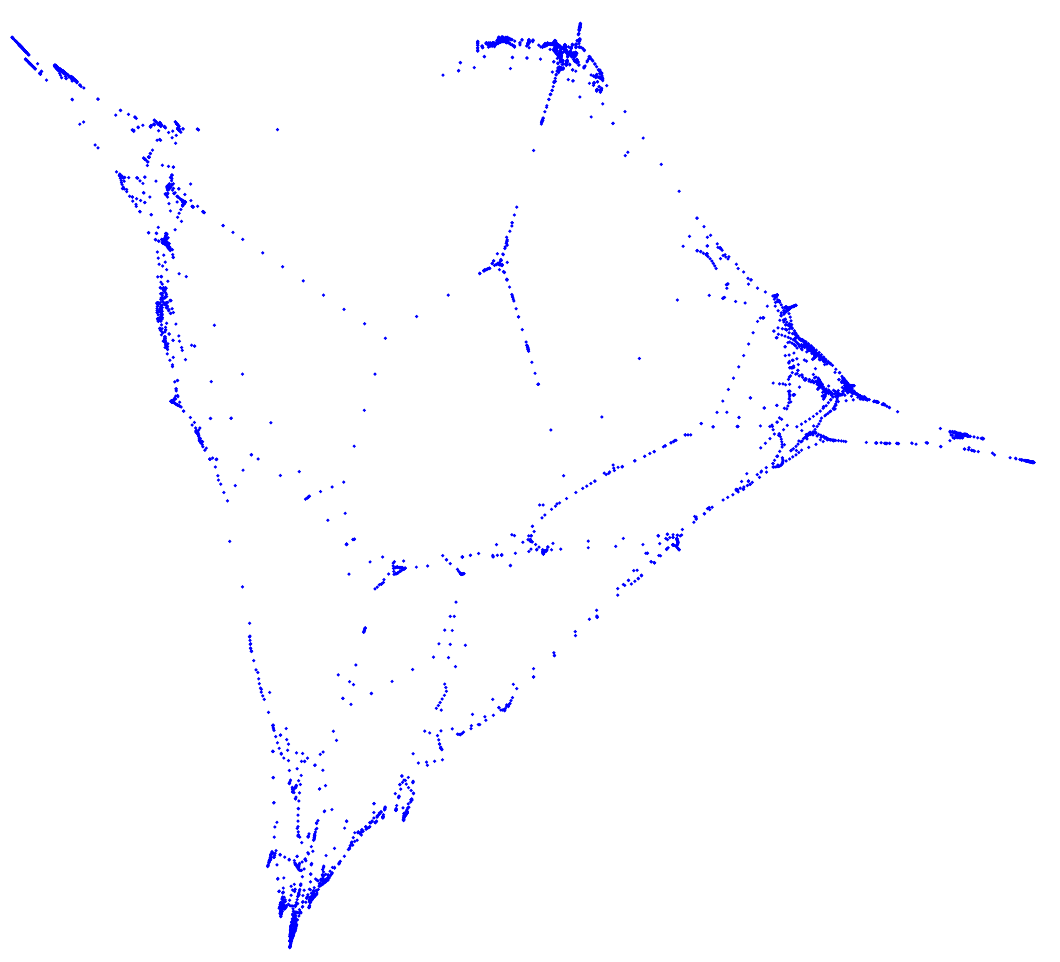}
    \label{fig:drawing-uni:powergrid}
  }
  \caption{
    Spectral graph drawings of two unipartite
    networks. \subref{fig:drawing-uni:zachary}~Zachary's karate club,
    \subref{fig:drawing-uni:powergrid}~the United States power grid
    (drawn without the edges for clarity). 
    Both networks are drawn using the two eigenvectors of smallest
    nonzero eigenvalue of the Laplacian matrix $\mathbf L$. 
  }
  \label{fig:drawing-uni}
\end{figure}

As an example of this method, we show the spectral drawing of Zachary's
unipartite karate club network and of the United States power grid
network~\cite{b228} in Figure~\ref{fig:drawing-uni}, showing that the
method can be applied to both small and large networks.

\subsection{Spectral Drawing of Bipartite Graphs}
\begin{figure}
  \centering
  \subfigure[Subset of artist--genre network]{
    \includegraphics[width=\wTwo]{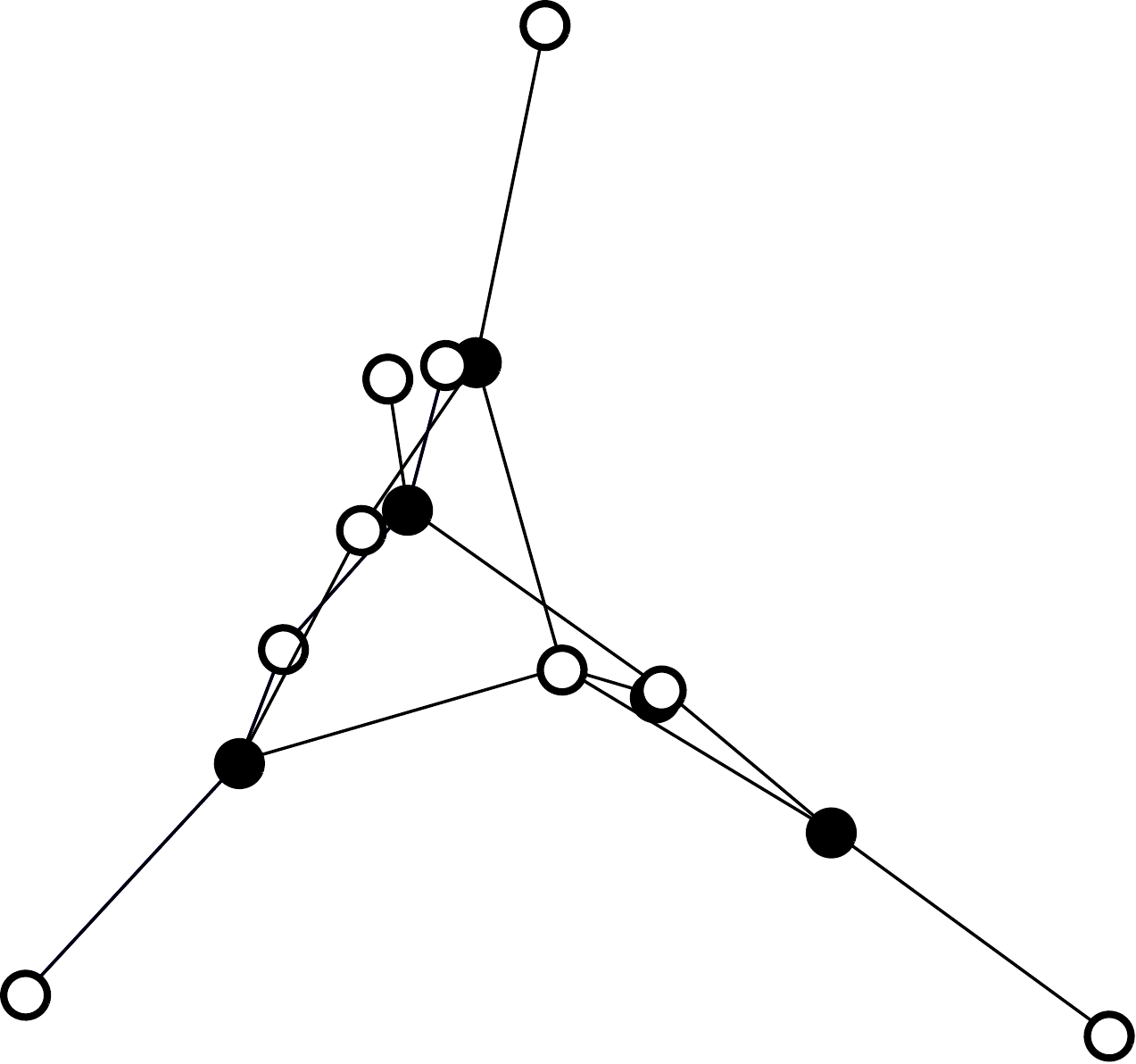}
    \label{fig:drawing-bip:small}
  }
  \subfigure[Full artist--genre network]{
    \includegraphics[width=\wTwo]{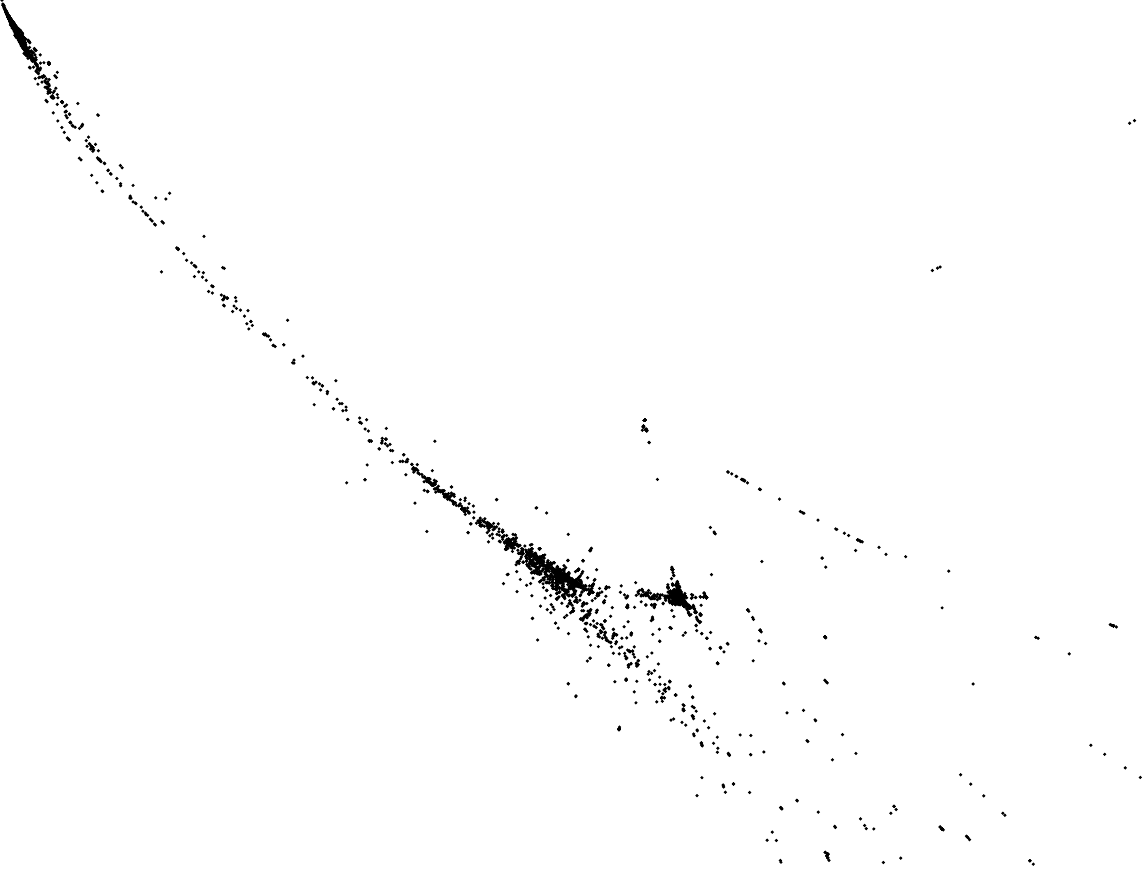}
    \label{fig:drawing-bip:large}
  }
  \caption{
    Spectral graph drawings of the bipartite artist--genre
    network. \subref{fig:drawing-bip:small}~small example subset,
    \subref{fig:drawing-bip:large}~full network (drawn without the edges
    for clarity).
    Both networks are drawn using the two eigenvectors of smallest
    nonzero eigenvalue of the Laplacian matrix $\mathbf L$. 
  }
  \label{fig:drawing-bip}
\end{figure}

The spectral graph drawing method introduced previously can be applied
to bipartite graphs, giving drawings where the nodes of the two types
are mixed in the plane.  An example for our small bipartite example
subset of the artist--genre network, as well as for the full
artist--genre network are shown in Figure~\ref{fig:drawing-bip}.

\begin{figure}
  \centering
  \includegraphics[width=\wTwo]{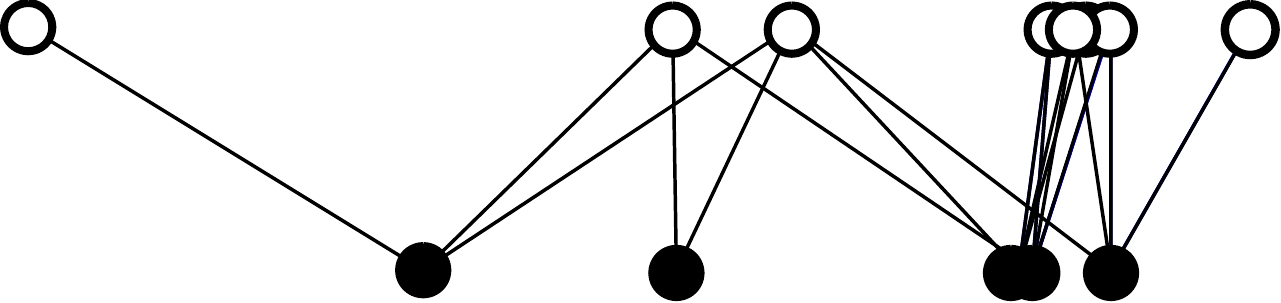}
  \caption{
    Drawing of the small bipartite artist--genre network that makes
    apparent the bipartite structure. Artists are shown as empty circles
    and genres as full circles.
  }
  \label{fig:drawing-bipline}
\end{figure}

Alternatively, we may want to emphasize the bipartite structure in the
network. Therefore, instead of using two eigenvectors of the Laplacian
matrix $\mathbf L$, we only use one. The coordinates for the other
dimension are then given by the bipartition:
\begin{align*}
  \mathbf X_{: 1} &= \mathbf x_2[\mathbf L] \\
  \mathbf X_{u2} &= \left\{ \begin{array}{ll}
    +1 & \text{ when } u \in V_1 \\
    -1 & \text{ when } u \in V_2
  \end{array} \right. 
\end{align*}                    
Note that this kind of bipartite graph drawing only makes sense when the
edges are shown. If, as we did with the large artist--genre and power
grid networks, edges are not shown, the drawing becomes uninformative
since it only consist of two lines of points. The drawing is shown for
the small bipartite artist--genre subset in
Figure~\ref{fig:drawing-bipline}.  As expected, the bipartivity of the
network becomes apparent from the graph drawing, and the placement of
the nodes on the two lines is such that clusters are apparent, for
instance the seven nodes at the right of the drawing.

In order to make apparent the bipartite structure of a network, other
methods exist beyond the two-line model, for instance drawing vertices
in two regions~\cite{b729} or using a radial layout~\cite{b730}.

\section{Link Prediction}
\label{sec:link-prediction}
The problem of \emph{link prediction} covers a set of related problems
that arise in various areas, which can all be formulated as the task of
predicting links in a network.  For instance, a social recommender
system such as the \emph{You may know these persons} feature on
Facebook\footnote{\href{https://www.facebook.com/find-friends/browser/}{www.facebook.com/find-friends/browser}}
can be implemented by predicting new friendship ties in the Facebook
social graph. Another application consists in finding researchers
similar to a given one based on their collaboration graph.  Yet another
application consists in predict the sign of edges in a signed social
network, i.e., a social network such as Slashdot
\cite{kunegis:slashdot-zoo} in which both \emph{friend} and \emph{foe}
ties are allowed.  All these applications are of the same form, in which
a graph as well as two nodes are given, and a score must be computed for
them.

The link prediction problem can be applied to bipartite networks,
too. For instance, recommending items on Facebook based on the
user--item \emph{like} graph amounts to predicting links in that graph.
A related task is collaborative filtering, i.e., the prediction of
ratings in the bipartite user--item rating graph.  This task differs
from the pure link prediction task, because edges are weighted by
ratings, and the goal is not to find edges, but to predict their weight.
Nonetheless, the task can be implemented by the same methods used for
link prediction.

The link prediction problem as defined here is very general, and plays a
central role in the areas of information retrieval (mainly through
word--document links), recommender systems (mainly through user--item
links) and other domains. Thus, the number of previously described
approaches is very large, and we only cover here a small representative
subset of them which have algebraic formulations: preferential
attachment, the common neighbors count, and various graph kernels.

A recommender can take several forms, not only by the type of network it
operates on, but also with regard to whether it finds nodes, edges or
computes probabilities.  In our experiments, we will use the most
general approach: returning a score when given a node pair.  Thus, all
link prediction algorithms for a graph $G=(V,E)$ will be formulated in
terms of a score function $f: |V|^2 \rightarrow \mathbb R$ that takes
two vertices as input, and returns a real number. We do not pose any
numerical constraint on the returned values; they could be
probabilities, counts, or even negative values. The only rule is that a
higher value denotes a higher probability of link formation.

\subsection{Preferential Attachment}
\emph{Preferential attachment} refers to the idea that when a link is
created in a network, it is more likely to attach to nodes that already
have a high degree. More precisely, the probability that a new link
attaches to a node $u$ can be modeled to be proportional to the degree
$d(u)$. When taken as the only link formulation rule, this leads to the
Barabási--Albert graph model.  As a solution to the link prediction
problem, preferential attachment can be used to derive the link
prediction function \cite{b439}
\begin{align*}
  f_{\mathrm{PA}}(u,v) &= d(u) d(v). 
\end{align*}
This is arguably the simplest nontrivial link prediction function
possible; it only takes vertex degrees into account, not the network
structure. In fact, if we chose any vertex $u$, compute this function
for the pairs $(u,v)$ for all $v \in |V|$, and then rank all vertices
$v$ by score, the result will be independent of the initial vertex
$u$. Thus, the preferential attachment link prediction function
corresponds to the \emph{most popular} recommender, i.e., a
non-personalized recommender function.  In the example of a friend
recommender for a social network, the preferential attachment function
always recommends the users with highest number of friends.

Even though it is rudimentary, the preferential attachment link
prediction function can give reasonable results in many applications,
and has the advantage to generalize trivially to bipartite networks: We
can use its definition directly in a bipartite network without any
modification.  Other, more complex link prediction functions however are
not that simple and will need adjustment for bipartite networks, since
they are defined in terms of the underlying network structure.

\subsection{Neighborhood-based Methods}
Beyond preferential attachment, the simplest link prediction functions
that are not independent on a source vertex are based on the principle
of triangle closing~\cite{b256}.  For instance, the common neighbors
count consists in counting the number of common neighbors between two
given nodes, and is based on the observation that an edge joining two
nodes with a high number of common neighbors will add many triangles to
the graph.  In the example of a friend recommender, this corresponds to
the \emph{friend of a friend} model, in which we recommend the users
with which we have the highest number of common friends, but who are not
our friends yet:
\begin{align*}
  f_{\mathrm{CN}}(u,v) &= |\{ w \in V \mid u \sim w, v \sim w \}|
\end{align*}
While this link prediction function takes into account the underlying
network structure, it is useless for bipartite networks. In fact, new
edges in a bipartite network always connect two vertices from different
partitions, and cannot have common neighbors.  In other words, triangle
closing does not apply to bipartite networks, since bipartite networks
do not contain triangles.  However, since the common neighbor count is
equal to the number of paths of length two between two nodes, we can
switch to using the number of paths of length three as a bipartite link
prediction function, which respects the bipartite network structure:
\begin{align*}
  f_{\mathrm{P3}}(u,v) &= |\{ w, x \in V \mid u \sim w, w \sim x, x \sim
  v \}|
\end{align*}
As shown in Section~\ref{sec:algebraic}, this can be expressed using the
biadjacency matrix as
$f_{\mathrm{P3}}(u,v)= [\mathbf B \mathbf B^{\mathrm T} \mathbf
  B]_{uv}$. 

A certain number of other link prediction functions are similar to the
common neighbors count, but use normalization or multiplicative factors
in various ways. These include the Jaccard index \cite{b256}, the cosine
similarity \cite{b676}, and the measure of Adamic and Adar~\cite{b475}.
The extension of these measures to bipartite graphs can be done in
various ways, depending on the generalization of the specific
normalization method.

\subsection{Graph Kernels}
As an example of a class of more advanced link prediction functions, we
review graph kernels. In a unipartite graph $G=(V,E)$, a graph kernel is
a function of two vertices $k: |V|^2 \rightarrow \mathbb R$ which is a
\emph{kernel}, i.e., the following properties hold:
\begin{itemize}
\item It is symmetric: $k(u,v) = k(v,u)$
\item It is positive semidefinite: For any vector $\mathbf x \in \mathbb
  R^{|V|}$, $k$'s bilinear form is nonnegative, i.e., $\sum_{u,v\in V}
  \mathbf x_u k(u,v) \mathbf x_v \geq 0$. 
\end{itemize}
A graph kernel thus has the same form as a link prediction function, but
with the additional condition of being symmetric and
positive-semidefinite.  Graph kernels can be defined algebraically using
the adjacency matrix and the Laplacian matrix, as in the examples given
in Table~\ref{tab:kernels}.

\begin{table}
  \caption{
    Examples of graph kernels that apply to unipartite networks. 
  }
  \centering
  \begin{tabular}{ll}
    \toprule
    \textbf{Name} & \textbf{Expression} \\
    \midrule
    Neumann kernel \cite{b263} &   $k_{\mathrm{NEU}}(u,v) = [(\mathbf I - \alpha \mathbf A)^{-1}]_{uv}$ \\
    Exponential kernel \cite{b156} &  $k_{\mathrm{EXP}}(u,v) = [e^{\alpha \mathbf A}]_{uv}$ \\
    Commute-time kernel \cite{b107} & $k_{\mathrm{COM}}(u,v) = [\mathbf L^+]_{uv}$ \\
    Heat diffusion kernel \cite{b137} & $k_{\mathrm{HEAT}}(u,v) = [e^{- \alpha \mathbf L}]_{uv}$ \\
    \bottomrule
  \end{tabular}
  \label{tab:kernels}
\end{table}

These graph kernels can be generalized to bipartite graphs. As an
example, the exponential kernel can be generalized to the hyperbolic
sine pseudokernel, and the special bipartite network structure can be
exploited to compute it.  The exponential kernel can be written as a
power sum:
\begin{align*}
  e^{\alpha \mathbf A} &= \mathbf I + \alpha \mathbf A + \frac {\alpha^2}2
  \mathbf A^2 + \frac {\alpha^3}6 \mathbf A^3 + \dotsb 
\end{align*}
In this form, it is clear that the exponential of the adjacency matrix
equals a sum over all paths between two nodes, weighted by a decreasing
function $\alpha^k / k!$ of the path length $k$. Thus, the exponential
function enjoys two useful properties as a link prediction function:
\begin{itemize}
\item It is higher when there are more paths connecting two nodes.
\item It is lower when the connecting paths are long. 
\end{itemize}
These two properties are very useful for a unipartite link prediction
function. For a bipartite link prediction function, the second property
is counterproductive. Instead, only paths of odd length should be
counted, and paths of even length have weighted zero.  This is because a
new edge in a bipartite network can only connect two nodes from
different partitions in the network, and thus pre-existing paths
connecting them must have odd length.  Thus, the corresponding graph
kernel for bipartite graphs is given by a reduction of the matrix
exponential to only its odd terms; this is exactly the matrix hyperbolic
sine~\cite{kunegis:hyperbolic-sine}:
\begin{align*}
  k_{\mathrm{SINH}} &= \sinh(\alpha \mathbf A) = \alpha \mathbf A +
  \frac {\alpha^3}6 \mathbf A^3 + \frac {\alpha^5} {120} \mathbf A^5 + \dotsb 
\end{align*}
This function however is not a kernel, since it is not positive
semidefinite -- which is reflected by the fact the hyperbolic sine takes
on negative values, unlike the exponential. Therefore, the resulting
link prediction function is only a \emph{pseudokernel}, i.e., a kernel
without the positive-semidefinite property.

Both the exponential kernel and the hyperbolic sine pseudokernel can be
computed efficiently using matrix decompositions. From its expression as
a power sum, it can be seen that the matrix exponential can be written
with help of the eigenvalue decomposition $\mathbf A = \mathbf U \mathbf
\Lambda \mathbf U^{\mathrm T}$:
\begin{align*}
  e^{\alpha \mathbf A} &= \mathbf U e^{\alpha \mathbf \Lambda} \mathbf U^{\mathrm T}
\end{align*}
Analogously, the hyperbolic sine can be written with help of the
singular value decomposition of the biadjacency matrix $\mathbf
B = \mathbf U \mathbf \Sigma \mathbf V^{\mathrm T}$:
\begin{align*}
  \sinh(\alpha \mathbf A) &= \left[ \begin{array}{cc}
      \mathbf 0 & \mathbf U \sinh(\alpha \mathbf \Sigma) \mathbf V^{\mathrm T} \\
      \mathbf V \sinh(\alpha \mathbf \Sigma) \mathbf U^{\mathrm T} & \mathbf 0 
    \end{array} \right]
\end{align*}

Analogously, we can derive a bipartite version of the Neumann kernel.
The Neumann kernel is defined as
\begin{align*}
  (\mathbf I - \alpha \mathbf A)^{-1} &= \mathbf I + \alpha \mathbf A +
  \alpha^2 \mathbf A^2 + \dotsb
\end{align*}
and, similarly to the exponential kernel, it assigns decreasing weights
to increasing path lengths. 
By keeping only the odd terms we arrive at the odd Neumann
kernel~\cite{b263}:  
\begin{align*}
  k_{\mathrm{NEU}} &= \alpha \mathbf A + \alpha^3 \mathbf A^3 + \alpha^5
  \mathbf A^5 + \dotsb = \alpha \mathbf A (\mathbf I - \alpha^2 \mathbf A^2)^{-1} 
\end{align*}
The same strategy can be used for other graph kernels based on the
adjacency matrix, for instance those learned from historical
data~\cite{kunegis:spectral-transformation}.  These kernels can also be
applied to the normalized adjacency matrix $\mathbf N$ and to the graph
Laplacian $\mathbf L$~\cite{b112,b373,b727}.  

\subsection{Evaluation}
\begin{figure}
  \centering
  \begin{tabular}{c}
    \includegraphics[width=0.45\textwidth]{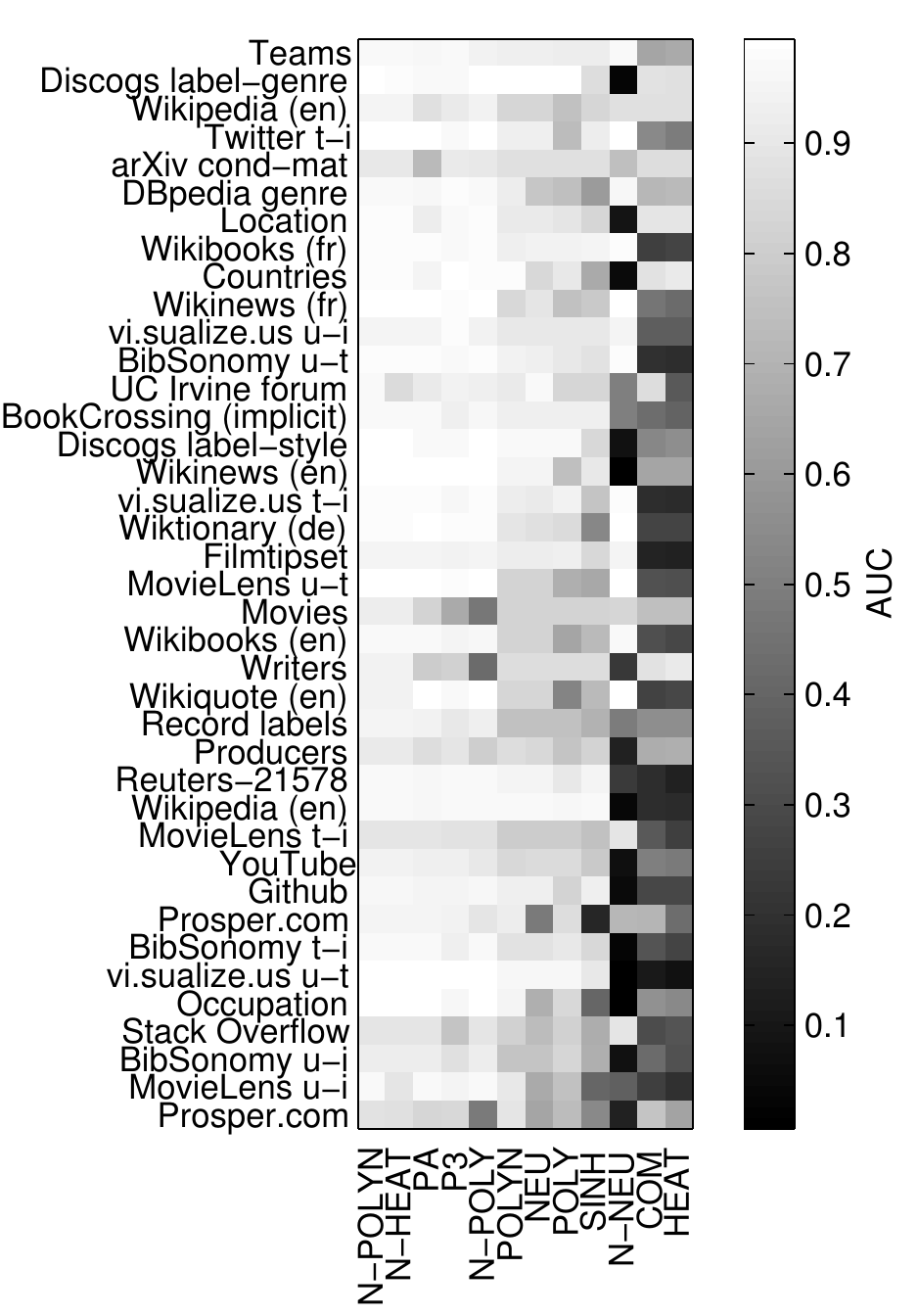}
  \end{tabular}
  \begin{tabular}{c}
    \includegraphics[width=0.2\textwidth]{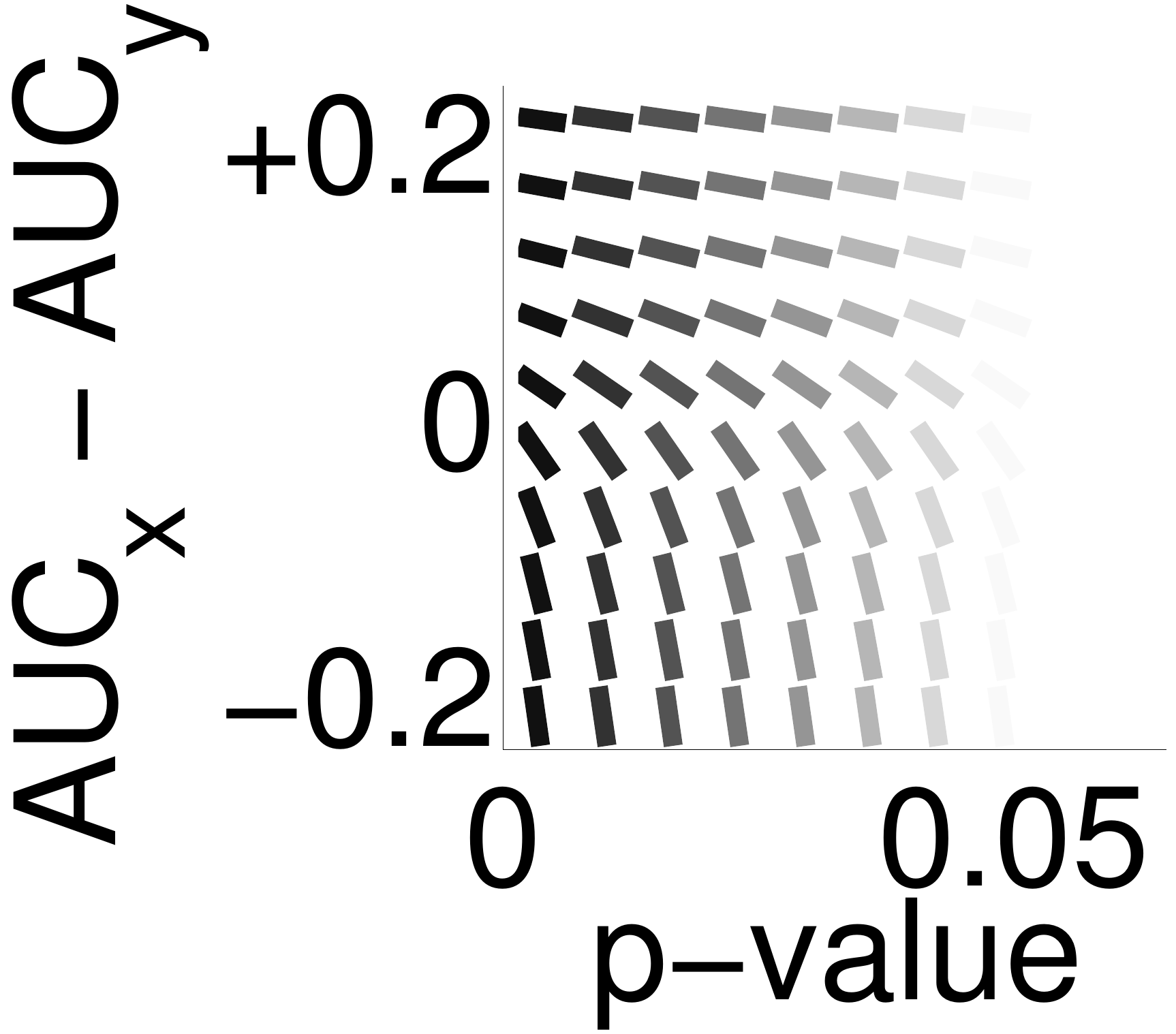} \\
    \includegraphics[width=0.45\textwidth]{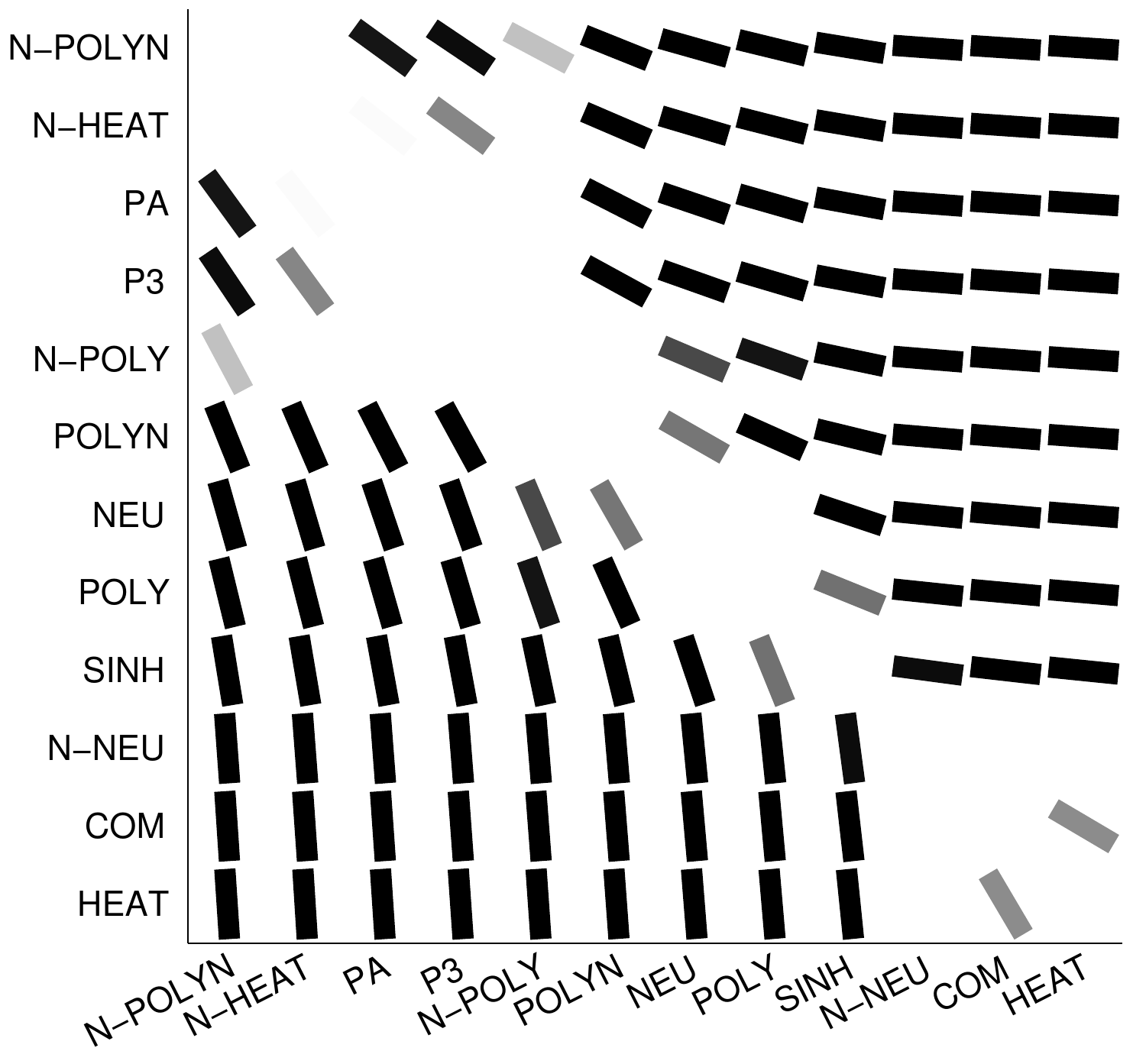}
  \end{tabular}
  \caption{
    \label{fig:link-prediction}
    Comparison of link prediction algorithms for bipartite networks.  
    Left:  the area under curve (AUC) for all methods and all networks.
    Right:  pairwise comparison of performances, showing the p-value of
    the comparison as well as the average difference in AUC values. 
  }
\end{figure}

We perform an evaluation of the described link prediction algorithms on
bipartite networks from the Koblenz Network Collection. The link
prediction methods we used are summarized in Table~\ref{tab:methods},
and the list of networks on which we make our experiments is given in
Appendix~\ref{sec:networks}.  We use the following methodology: For each
network, we divide the set of edges into a training set (75\% of edges)
and a test set (25\% of edges). Additionally, a zero test set of
unconnected node pairs is created, having the same size as the test set.
When edge arrival times are known for one network, the split is such
that all training edges come before test edges. We then use the various
methods shown in Table~\ref{tab:methods} to compute link prediction
scores for all node pairs in the test set and zero test set.  The scores
are then used to compute the area under curve (AUC) as a measure of
precision for each method~\cite{b366}.  AUC values range from zero to
one, with higher values denoting better ranking.  The results are shown
in Figure~\ref{fig:link-prediction}.

\begin{table}
  \caption{
    \label{tab:methods}
    List of bipartite link prediction methods used in our experiments. 
  }
  \centering
  \scalebox{0.95}{
    \begin{tabular}{l p{5.5cm} l}
      \toprule
      & \textbf{Name} & \textbf{Expression} \\
      \midrule
      PA & Preferential attachment & $d(u) d(v)$ \\
      P3 & Paths of length three & $|\{ u \sim x \sim y \sim v \}|$ \\
      \midrule
      POLY & Odd polynomial of the adjacency matrix &
      $P_{\mathrm{o}}(\mathbf A)$ \\
      POLYN & Odd nonnegative polynomial of the adjacency matrix &
      $P_{\mathrm{on}}(\mathbf A)$ \\
      NEU & Odd Neumann kernel & $\alpha \mathbf A (\mathbf I - \alpha^2 \mathbf A^2)^{-1}$ \\
      SINH & Hyperbolic sine & $\sinh(\alpha \mathbf A)$ \\
      \midrule
      N-POLY & Odd polynomial of the normalized adjacency matrix &
      $P_{\mathrm o}(\mathbf N)$ \\
      N-POLYN & Odd nonnegative polynomial of the normalized adjacency
      matrix & $P_{\mathrm{on}}(\mathbf N)$  \\
      N-NEU & Normalized odd Neumann kernel & $\alpha \mathbf N (\mathbf I - \alpha^2 \mathbf N^2)^{-1}$ \\ 
      N-HEAT & Normalized heat diffusion kernel & $\sinh(\alpha \mathbf
      N)$ \\
      \midrule
      COM & Commute-time kernel & $\mathbf L^+$ \\
      HEAT & Heat diffusion kernel & $ \exp(-\alpha \mathbf L)$ \\
      \bottomrule
    \end{tabular}
  }
\end{table}

Our experiments show that not all link prediction methods generalize
well to bipartite networks.  First, we see that the two methods based on
the Laplacian matrix, the heat diffusion kernel and the commute-time
kernel, perform very badly.  The best results are in fact achieved by
two functions of the normalized adjacency matrix:  the polynomial and
the normalized heat diffusion kernel.  The normal Neumann kernel
however does not perform well.  In a similar vein to unipartite
networks, we must conclude that there is no generally best method, and
that the performance can vary from dataset to dataset as it does from
method to method. 

\section{Other Graph Types}
\label{sec:other}
Graphs as defined in the previous sections of this article are generally
called simple graphs, as they allow only single edges between two
vertices.  Their edges are undirected and unweighted.  The archetypal
example of a network modeled as a simple graph is a social network:
persons connected by friendships.  This definition can be extended to
include multiple edges between a single vertex pair, weighted edges,
signed edges, and directed edges.  Of these extensions, all can be
applied to bipartite graphs.  In this section, we briefly review two of
these extensions and their implications for algebraic graph theory:
weighted graphs and directed graphs.

\paragraph{Edge Weights}
A common feature of graphs are edge weights, which come in multiple
forms.  For one, edges can be directly weighted by a real number, such
as in the trust network of the Advogato online community, in which three
levels of trust exist, represented by the numbers 0.6, 0.8 and
1.0~\cite{b697}.  Another type of weight is given by signed networks, in
which positive and negative edges exist.  An example is the Slashdot Zoo
social network, in which people are related by \emph{friend} and
\emph{foe} links \cite{kunegis:slashdot-zoo}.  Finally, networks may
allow multiple parallel edges, in which case we may model the number of
parallel edges as an integer weight.  An example is an email network,
with each edge representing an email between two persons.  All these
types of weightings can be applied to bipartite networks directly,
giving a weighted biadjacency matrix $\mathbf B$.  This matrix can then
be used for any algebraic graph analysis method trivially.

\paragraph{Directed Graphs}
In directed graphs, each edge has an orientation.  Instead of being
defined as the unordered set $\{u,v\}$, an edge in a directed network is
defined as the ordered pair $(u,v)$. Edges in a directed graph are also
called \emph{arcs}.  An example for a directed graph is the Twitter
follower graph, consisting of Twitter users and their follow
relationships~\cite{b545}.  The fact that user $u$ follows user $v$ can
be represented by the directed edge $(u,v)\in E$.  Arguably, more
directed than undirected unipartite networks exist; in the Koblenz
Network Collection for instance, 62\% of all unipartite networks are
directed~\cite{konect}.  On the other hand, directed bipartite graphs
are very rare.  They occur for instance in Petri nets, where the node
types are places and transitions. Out of 189 networks in the Koblenz
Network Collection as of 2014, not a single one is directed and
bipartite.

\begin{figure}[h]
  \centering
  \includegraphics[height=0.20\textwidth]{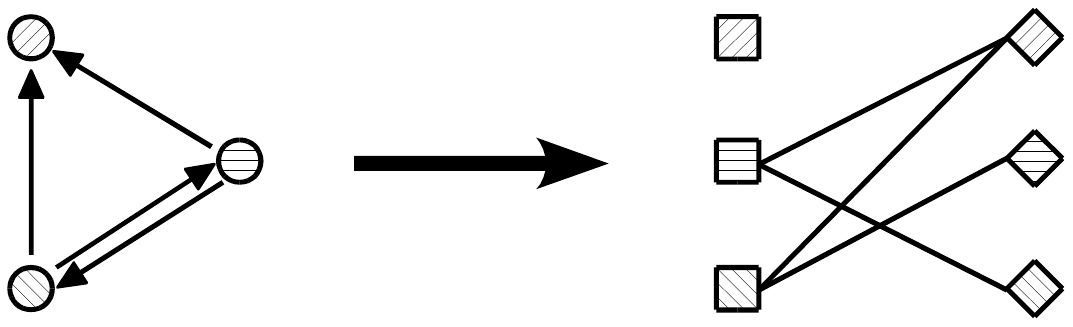}
  \caption{
    \label{fig:cover}
    Illustration of the bipartite double cover, which generates an
    undirected bipartite graph from any directed unipartite graph. 
  }
\end{figure}

Directed graphs can be transformed into bipartite graphs, using the
bipartite double cover~\cite{b667}. This gives a bipartite graph in
which each edge corresponds to an original directed edge, and the set of
vertices is doubled, as shown in Figure~\ref{fig:cover}.
Formally, the directed graph $G=(V,E)$ has the bipartite double cover
\begin{align*}
  H &= (V\times \{1,2\}, \{ \{(u, 1), (v,2)\} \mid (u,v) \in E \}). 
\end{align*}
This mapping does not preserve the complete directed structure, because
directed paths of the form $u \rightarrow v \rightarrow w$ are not
reflected as paths in the bipartite double cover; only
\emph{alternating} paths are preserved, i.e., paths consisting of edges
in alternating orientations relative to the path such as $u \rightarrow
v \leftarrow w \rightarrow x$.  If $\mathbf B$ is the adjacency matrix
of a directed graph, then $\mathbf B$ is also the biadjacency matrix of
its double cover, whose adjacency matrix is
\begin{align*}
  \left[ \begin{array}{cc} 
      \mathbf 0 & \mathbf B \\
      \mathbf B^{\mathrm T} & \mathbf 0 
    \end{array} \right] .
\end{align*}
An example in which these properties of the bipartite double cover are
exploited are rating networks of online dating
websites~\cite{kunegis:split-complex-dating}.

\section*{Conclusion}
The topic of algebraic graph theory for bipartite networks is vast, and,
as we have seen, specific analysis methods can be defined for many
applications. In most cases, analysing a bipartite network does not just
mean applying regular network analysis methods to a network that happens
to the bipartite. Instead, specialized algebraic methods exist, that are
both more efficient and better suited to the bipartite structure.  A
general pattern that we have exploited is the replacement of the square
and symmetric adjacency matrix $\mathbf A$ with the rectangular
biadjacency matrix $\mathbf B$.  This construction leads to changes in
its applications.  For instance, the eigenvalue decomposition must be
replaced by the singular value decomposition, which naturally maps to
bipartite graphs, and the exponential kernel must be replaced by the
hyperbolic sine pseudokernel.  We also note that the existence of a
match between the singular value decomposition and bipartite graphs is
not trivial, in that there are examples of graphs with special structure
that do not allow such a mapping: Directed networks for instance are
represented naturally by asymmetric and square adjacency matrices, for
which no extension of the eigenvalue decomposition is satisfactory, as
shown in \cite{kunegis:directed-decomposition}: the asymmetric
eigenvalue decomposition is not defined in the general case and discards
information about when it is defined, the singular value decomposition
of the asymmetric adjacency matrix implicitly applies to the bipartite
double cover and thus loses information, and an alternative
decomposition (DEDICOM) is computationally complex and non-unique.

The concept of bipartite networks itself can be generalized to
$k$-partite networks. There are two different ways of achieving this:
First, by letting each edge connect one vertex of every of the $k$
vertex sets, resulting in a $k$-regular hypergraph.  This approach is
taken for instance in the modeling of broad folksonomies, in which each
tag assignment consists of a user--item--tag triple~\cite{b851}.  The
second generalization of bipartite networks also allows any of $k$
different vertex types, and also $l$ different edge types, each edge
type connecting two specific vertex types (which may be the same)
\cite{b604}. This approach is taken in semantic networks, where each
edge type then corresponds to one semantic predicate.

\appendix
\section{Appendix:  Networks}
\label{sec:networks}
The list of networks used in the experiments in this paper is given in
Table~\ref{tab:networks}.  
All networks are available online with the Koblenz Network
Collection\footnote{\href{http://konect.uni-koblenz.de/}{konect.uni-koblenz.de}}~\cite{konect}.  

\begin{table}
  \caption{
    \label{tab:networks}
    The bipartite networks used in this article. 
  }
  \makebox[\textwidth]{
    \scalebox{0.92}{
      \input{tex-bipartivity/networks}
    }
  }
\end{table}

\bibliographystyle{abbrv}

\bibliography{all}

\end{document}

%% file: tex-bipartivity/scattercorr.anticonflict.nonbip.tex
$\rho = 0.06$

%% file: tex-bipartivity/scattercorr.anticonflict.nonbipn.tex
$\rho = 0.52^{***}$

%% file: tex-bipartivity/scattercorr.anticonflict.oddcycles.tex
$\rho = -0.08$

%% file: tex-bipartivity/scattercorr.nonbip.nonbipn.tex
$\rho = 0.14$

%% file: tex-bipartivity/scattercorr.nonbip.oddcycles.tex
$\rho = 0.32^{**}$

%% file: tex-bipartivity/scattercorr.nonbipn.oddcycles.tex
$\rho = 0.11$

%% file: tex-bipartivity/networks.tex
\begin{tabular}{lrrll}
\toprule
\textbf{Network} & $\boldsymbol{|V|}$ & $\boldsymbol{|E|}$ & \textbf{Nodes} & \textbf{Edges} \\ 
\midrule
BibSonomy t-i & $972{,}120$ & $2{,}555{,}080$ & Tag +  Publication & Assignment \\ 
BibSonomy u-i & $773{,}241$ & $2{,}555{,}080$ & User +  Publication & Tag assignment \\ 
BibSonomy u-t & $210{,}467$ & $2{,}555{,}080$ & User +  Tag & Assignment \\ 
BookCrossing (implicit) & $445{,}801$ & $1{,}149{,}739$ & User + Book & Rating \\ 
CiteULike u-t & $175{,}992$ & $2{,}411{,}819$ & User +  Tag & Assignment \\ 
Countries & $592{,}414$ & $637{,}134$ & Entity +  Country & Affiliation \\ 
DBpedia genre & $266{,}836$ & $470{,}223$ & Work +  Genre & Style \\ 
Location & $225{,}498$ & $296{,}475$ & Entity +  Place & Location \\ 
Occupation & $230{,}167$ & $253{,}199$ & Person +  Occupation & Association \\ 
Producers & $187{,}721$ & $208{,}999$ & Producer +  Work & Association \\ 
Record labels & $186{,}971$ & $236{,}766$ & Artist +  Record & Membership \\ 
Movies & $157{,}199$ & $281{,}905$ & Movie +  Actor & Starring \\ 
Teams & $937{,}011$ & $1{,}492{,}503$ & Athlete +  Team & Membership \\ 
Writers & $135{,}575$ & $147{,}512$ & Writer +  Work & Authorship \\ 
Discogs label–genre & $270{,}786$ & $4{,}147{,}665$ & Label + Genre & Feature \\ 
Discogs label–style & $244{,}147$ & $5{,}255{,}950$ & Label + Style & Feature \\ 
Wiktionary (de) & $151{,}982$ & $1{,}229{,}501$ & User +  Article & Edit \\ 
Wikibooks (en) & $167{,}525$ & $1{,}164{,}576$ & User +  Article & Edit \\ 
Wikinews (en) & $173{,}772$ & $901{,}416$ & User +  Article & Edit \\ 
Wikiquote (en) & $116{,}363$ & $549{,}210$ & User +  Article & Edit \\ 
Wikibooks (fr) & $30{,}997$ & $201{,}727$ & User +  Article & Edit \\ 
Wikinews (fr) & $26{,}546$ & $193{,}618$ & User +  Article & Edit \\ 
Filmtipset & $75{,}360$ & $1{,}266{,}753$ & User +  Movie & Comment \\ 
Github & $177{,}386$ & $440{,}237$ & User +  Project & Membership \\ 
Wikipedia (en) & $276{,}739$ & $2{,}941{,}902$ & Article +  Word & Inclusion \\ 
Reuters-21578 & $60{,}234$ & $978{,}446$ & Article +  Word & Inclusion \\ 
MovieLens t-i & $24{,}129$ & $95{,}580$ & Tag +  Movie & Tag assignment \\ 
MovieLens u-i & $11{,}610$ & $95{,}580$ & User +  Movie & Tag assignment \\ 
MovieLens u-t & $20{,}537$ & $95{,}580$ & User +  Tag & Assignment \\ 
Twitter t-i & $1{,}502{,}611$ & $2{,}635{,}885$ & Hashtag +  URL & Co-occurrence \\ 
arXiv cond-mat & $38{,}741$ & $58{,}595$ & Author + Paper & Authorship \\ 
Southern women & $32$ & $89$ & Woman + Event & Participation \\ 
UC Irvine forum & $1{,}421$ & $33{,}720$ & User + Forum & Post \\ 
vi.sualize.us t-i & $577{,}437$ & $2{,}298{,}816$ & Tag +  Picture & Assignment \\ 
vi.sualize.us u-i & $512{,}524$ & $2{,}298{,}816$ & User +  Picture & Tag assignment \\ 
vi.sualize.us u-t & $99{,}157$ & $2{,}298{,}816$ & User +  Tag & Assignment \\ 
Prosper.com & $7{,}595$ & $21{,}017$ & Member +  Group & Support \\ 
Prosper.com & $23{,}965$ & $35{,}377$ & Member +  Listing & Watch \\ 
Stack Overflow & $641{,}876$ & $1{,}302{,}439$ & User + Post & Favorite \\ 
Wikipedia (en) & $2{,}036{,}440$ & $3{,}795{,}796$ & Article +  Category & Inclusion \\ 
YouTube & $124{,}325$ & $293{,}360$ & User +  Group & Membership \\ 
\bottomrule
\end{tabular}